\documentclass[journal]{new-aiaa}
\usepackage[utf8]{inputenc}
\usepackage{textcomp}

\usepackage{graphicx}
\usepackage{amsmath}
\usepackage{tikz}
\usetikzlibrary{shapes}  
\usetikzlibrary{shapes.geometric}  
\usepackage{xcolor}
\definecolor{customgreen}{RGB}{34, 139, 34}  

\newcommand{\blackcirc}{
    \tikz{\node[draw=black, fill=black, circle, minimum size=0.5em] {};}
}
\usepackage{pifont}
\newcommand{\mystar}{\Pisymbol{pzd}{75}}
\usepackage[version=4]{mhchem}
\usepackage{siunitx}
\usepackage{longtable,tabularx}
\usepackage{subcaption}
\newcommand{\fig}{Fig.}
\newcommand{\figs}{Figs.}
\newcommand{\revision}[1]{\textcolor{black}{#1}}

\title{Model-Based Reinforcement Learning for Control of Strongly-Disturbed Unsteady Aerodynamic Flows}

\author{Zhecheng Liu\footnote{Ph.D. Student, Department of Mechanical and Aerospace Engineering; zliu163@ucla.edu}}
\affil{University of California, Los Angeles, Los Angeles, California 90095}

\author{Diederik Beckers\footnote{Postdoctoral Researcher, Graduate Aerospace Laboratories; beckers@caltech.edu}}
\affil{California Institute of Technology, Pasadena, California 91125}

\author{Jeff D. Eldredge\footnote{Professor, Department of Mechanical and Aerospace Engineering; jdeldre@ucla.edu. Associate Fellow AIAA.}}
\affil{University of California, Los Angeles, Los Angeles, California 90095}

\begin{document}

\maketitle

\begin{abstract}
The intrinsic high dimension of fluid dynamics is an inherent challenge to control of aerodynamic flows, and this is further complicated by a flow's nonlinear response to strong disturbances. Deep reinforcement learning, which takes advantage of the exploratory aspects of reinforcement learning (RL) and the rich nonlinearity of a deep neural network, provides a promising approach to discover feasible control strategies. However, the typical model-free approach to reinforcement learning requires a significant amount of interaction between the flow environment and the RL agent during training, and this high training cost impedes its development and application. In this work, we propose a model-based reinforcement learning (MBRL) approach by incorporating a novel reduced-order model as a surrogate for the full environment. The model consists of a physics-augmented autoencoder, which compresses high-dimensional CFD flow field snaphsots into a three-dimensional latent space, and a latent dynamics model that is trained to accurately predict the long-time dynamics of trajectories in the latent space in response to action sequences. \revision{The accuracy and robustness of the model are demonstrated in the scenario of a pitching airfoil within a highly disturbed environment. Additionally, an application to a vertical-axis wind turbine in a disturbance-free environment is discussed in the Appendix.} Based on the \revision{model trained in the pitching airfoil problem}, we realize an MBRL strategy to mitigate lift variation during gust-airfoil encounters. We demonstrate that the policy learned in the reduced-order environment translates to an effective control strategy in the full CFD environment.
\end{abstract}

\section*{Nomenclature}
{\renewcommand\arraystretch{1.0}
\noindent\begin{longtable*}{@{}l @{\quad=\quad} l@{}}
$A_k$ & action from agent to environment at time step k\\ 
$C_l$& lift coefficient \\
$C_p$ & power coefficient \\
$c$ & chord length of a flat plate \\
$\mathbf{c}_k$ & cell state for a long short-term memory (LSTM) cell at time step k \\
$D_x$ & amplitude of Gaussian forcing field in horizontal direction \\
$D_y$ & amplitude of Gaussian forcing field in vertical direction \\
$\mathbf{F}$ & a forcing field varying spatially and temporally with a Gaussian form\\
$\mathcal{F}_{e,d,m}$ & encoder, decoder, and multi-layer perceptrons (MLP) component of the physics-augmented autoencoder (PA-AE) \\
$\mathbf{h}_k$ & hidden state for a LSTM cell at time step k\\ 
$\mathcal{J}$  & loss function for PA-AE \\
$O_k$ & observation from environment to agent at time step k\\ 
$\mathbf{P}$ & auxillary parameters for latent dynamics model (LDM) \\
$R_k$ & reward function from environment to agent at time step k\\
$R$ & distance between a rotating flat wing and the connected hub\\
$\mathbf{S}_k$ & flow state at time step k \\ 
$\mathbf{\tilde{S}}_k$ & reconstructed flow state from PA-AE at time step k \\ 
$\mathbf{\hat{S}}_k$ & predicted flow state from LDM with decoder at time step k \\ 
$T_p$ & time for a complete rotation cycle of a flat wing \\
$t_0$ & temporal mean of the Gaussian forcing field \\ 
$U_{\infty}$ & horizontal free stream velocity  \\
$\dot{W}$ & net power delivered to a body from surrounding fluids \\
$x_0$ & mean x-coordinate of the Gaussian forcing field \\ 
$y_0$ & mean y-coordinate of the Gaussian forcing field \\
$\mathbf{x}_i$ & the $i^{th} ~ (i\geq 1)$ element in the input sequence for a stacked LSTM \\
$\mathbf{y}_i$ & the $i^{th} ~ (i\geq 1)$ element in the output sequence for a stacked LSTM \\
$\alpha$ & pitching angel for a flat wing \\
$\dot{\alpha}$ & angular velocity for a flat wing \\
$\ddot{\alpha}$ & angular acceleration for a flat wing \\
$\Delta t$ & flow solver time step size\\
$\boldsymbol{\gamma}$ & latent variables \\
$\boldsymbol{\hat{\gamma}}$ & predicted latent variables from LDM \\
$\theta$ & rotating angle for a flat wing\\
$\rho$ & density\\
$\sigma_t$ & temporal width of the Gaussian forcing field \\
$\sigma_x$ & horizontal width of the Gaussian forcing field \\
$\sigma_y$ & vertical width of the Gaussian forcing field \\
$\Omega$ & rotating velocity for a flat wing\\
\multicolumn{2}{@{}l}{Subscripts}\\
k & discrete time step index representing the pitching action steps\\
\end{longtable*}}

\section{Introduction}\label{sec:introduction}
\lettrine{T}{he} control of unsteady flows plays a critical role in enhancing the performance, efficiency, and safety of various systems across a multitude of fields, including energy, transportation, and biomedical science. In most of these fields, there is an inherent challenge due to the strong nonlinearity generated from the interactions between the system, e.g., an air vehicle, and large-amplitude external flow disturbances and control inputs, like pitching, synthetic jets, and control surface actuation. Additionally, the large parameter space \cite{annurev:/content/journals/10.1146/annurev-fluid-031621-085520} of some flow disturbances, such as gusts, further complicates flow control efforts.

Flow control can be active or passive, but due to the fact that active flow control is more adaptable to different flow conditions, it usually has stronger performance than passive methods \revision{for achieving control objectives in aerodynamics} \cite{Ozkan2022}. Based on different control objectives, control input types, and measurement availability, notable examples of applying a control theoretic strategy to unsteady aerodynamics problems include utilizing a linear quadratic regulator to stabilize the unstable steady state of flow past a flat plate by applying a localized body force near the leading edge as actuation \cite{Ahuja2009FeedbackCO}, mitigating the lift transient for a flat plate during transverse wing-gust encounters through \revision{a} pitching \revision{controller designed based on unsteady aerodynamic models and proportional feedback control} \cite{Sedky_Gementzopoulos_Andreu-Angulo_Lagor_Jones_2022}, suppressing the lift disturbances on a low-aspect-ratio semicircular wing in a longitudinally gusting ﬂow through variable-pressure pulsed-blowing actuation with robust controllers that are synthesized using a mixed-sensitivity loop-shaping approach \cite{doi:10.2514/1.J050954}, and many others.

However, the practical application of classical linear approaches can be limited by strong nonlinear effects generated from large-amplitude disturbances or a rapid succession of actuation. Fortunately, the development of machine learning and data science potentially provides us with tools to better model and control these complex flow problems. \revision{In particular, reinforcement learning (RL) has been strengthened by advancements in deep learning, which enables more expressive nonlinear function approximation and enhances RL's ability to explore effective policies. This integration has enabled RL to tackle many systems with similar complexity successfully.} Examples include exploring an effective swimming strategy \cite{Novati_2017,Verma2018EfficientCS}, reducing drag of a circular cylinder \cite{doi:10.1073/pnas.2004939117,Rabault_Kuchta_Jensen_Réglade_Cerardi_2019,act11120359}, and minimizing the lift variations for a flat plate in a highly disturbed environment \cite{beckers2024deepreinforcementlearningairfoil}. However, in its usual form, RL is model free, and learns an effective control policy via a significant number of interactions directly between the RL agent and the full environment. This can be prohibitively expensive in a fluid dynamics context, in which the environment consists of a flow experiment or a CFD simulation. In a CFD simulation, the high cost comes from the intrinsic high dimensionality of a large number of grid points necessary to obtain a well resolved solution. Therefore, it is necessary to reduce the dimensionality of the environment in order to overcome the high cost of RL training. By constructing a low-order dynamical model that suitably approximates the agent's interaction with the full environment, we could do the RL training more economically in this surrogate model environment, then, when desired, evaluate the policy in the full environment. We take this approach in the current work, \revision{and classify it as a form of} model-based reinforcement learning (MBRL). We should clarify that the definition of MBRL is very broad. In many MBRL algorithms the model is internal to the RL agent; it is used for purposes of planning and is improved through the agent's ongoing interaction with the full environment during policy training \cite{moerland2023model}. However, in this study, we separate these tasks: the model is learned prior to policy training and, during this training, the RL agent interacts exclusively with the reduced-order environment provided by this model.

\begin{figure}[hbt!]
\centering
\includegraphics[width=1.0\textwidth]{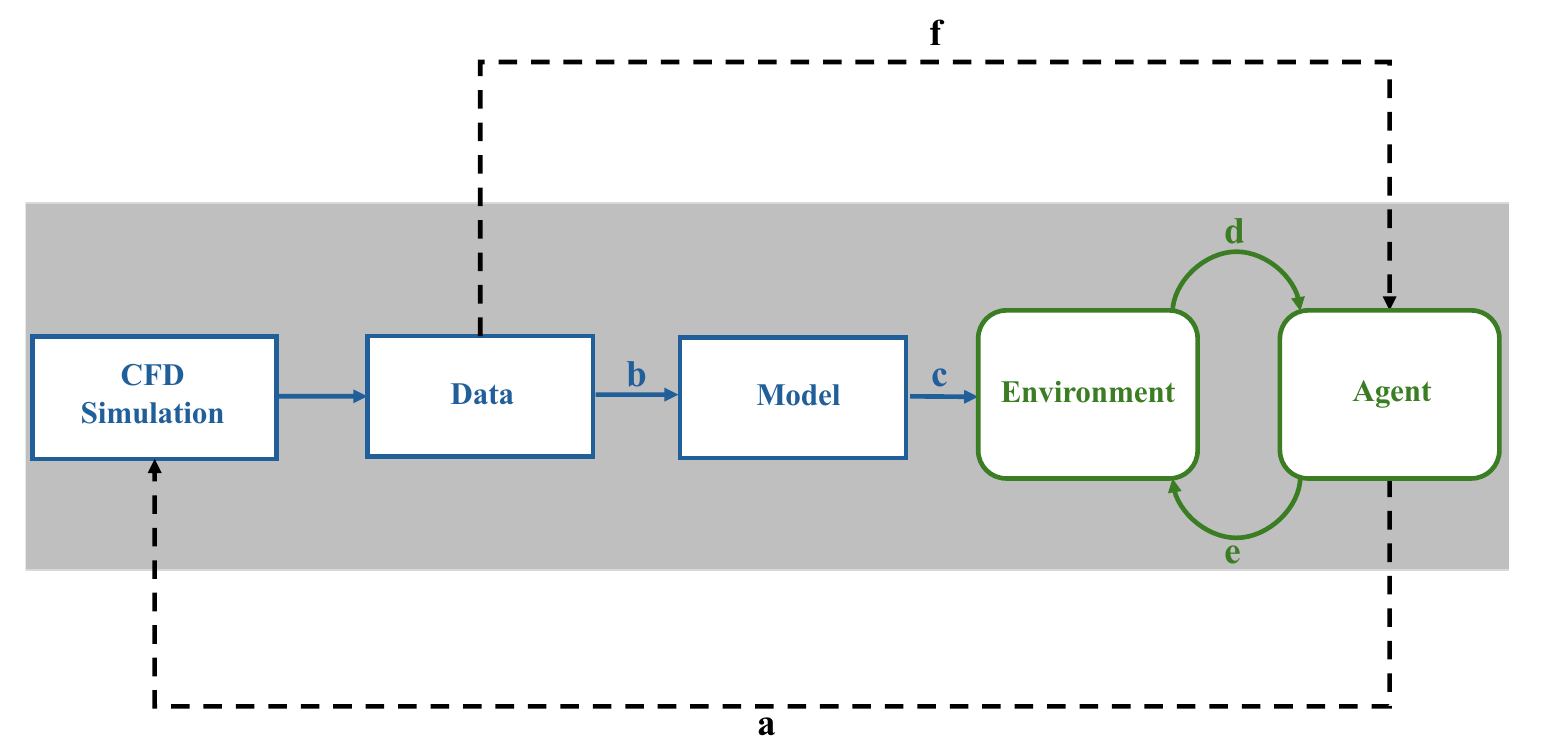}
\caption{The comparison between the implemented MBRL structure and the typical model-free RL structure. The MBRL structure is shown as solid lines, and the typical model-free structure is shown as dashed lines.}
\label{MBRL_Overview}
\end{figure}

A schematic figure of our implemented MBRL structure, contrasted with the typical model-free RL structure, is depicted in \fig~\ref{MBRL_Overview}. The specific connections implemented in this study are highlighted in the shaded block. Initially, we execute CFD simulations using a random policy to gather data. Subsequently, the reduced order model is learned from this collected data, as indicated by arrow b. This learned model is then inserted into the RL environment, as shown by arrow c, acting as a surrogate for the full environment behavior. Therefore, the RL agent is trained through interactions with the model environment, indicated by arrows d and e. In contrast, in a model-free application of RL, the agent would converge towards its policy by iteratively calling the CFD simulation directly and collecting its resulting data, as in arrows a and f. We have to clarify that the MBRL structure we study here is only one possibility, and for a thorough review of \revision{additional} MBRL structures, such as those incorporating planning, we refer the reader to Moerland et al.~\cite{moerland2023model}.

Since we will use a low dimensional model, described in \fig~\ref{MBRL_Overview}, to replace the CFD simulation, it is essential to first train a model for state abstraction before we train another model to learn the dynamics in the low dimensional space. Many advances in state abstraction benefit from representation learning, which is a crucial topic in RL and control \cite{LESORT2018379}. A good representation is important for dynamics model learning and also for policy and value function learning \cite{moerland2023model}. An early application of representation learning in the fluid dynamics community is proper orthogonal decomposition (POD) \cite{Berkooz1993ThePO,2017_Taira}, which finds a low-dimensional representation of the flow state through a linear decomposition into a small number of primary spatial modes extracted from a sequence of snapshots of a single flow. This representation is optimal for the flow on which the modes were derived, but not on other flows, e.g, with different configuration of the geometry or different disturbances or actuation. For compressing a variety of similar flows, researchers in fluid dynamics have recently drawn from tools in deep learning. In particular, the autoencoder \cite{doi:10.1126/science.1127647,Murata_Fukami_Fukagata_2020,Fukami2023} has recently been 
employed to compress high dimensional flow data to an even lower dimensional space than POD. The approach can be effective if the training is regularized with physics. For example, Fukami and Taira \cite{Fukami2023} found that both principal component analysis and a naive autoencoder struggled to meaningfully compress the flow states from a wide number of cases of two-dimensional flow past a wing at various angles of attack in extreme aerodynamic environments. By augmenting the decoder with an additional network that predicts lift, they showed that this lift-augmented autoencoder not only compresses extreme aerodynamic data into a three-dimensional latent space but also identifies universal flow features. We will follow this approach in the present work to compress (and expand) the full flow state, generalizing the augmentation for predicting lift to other physical variables \revision{(see Appendix)}, and refer to it as the physics-augmented autoencoder (PA-AE).

Generally, a dynamics model for RL applications consists of one of three types \cite{moerland2023model}: a forward model predicting the next state given a current state and a chosen action, a backward model predicting which states are the possible precursors of the current state, or an inverse model predicting which action is needed to get from current state to the next state. However, in most cases, we are interested in the forward model for MBRL, and in this work, we wish to learn a forward model for the low-dimensional latent-space dynamics for a wide spectrum of disturbed and actuated aerodynamic flows. We refer to this throughout the paper as the latent dynamics model (LDM). Dynamic mode decomposition (DMD) \cite{SCHMID_2010} is widely used in fluid dynamics to decompose the dynamics for systems that exhibit quasiperiodic behaviors by extracting a set of modes that isolate the dominant frequencies and intrinsic dynamic behavior of the system. However, DMD generally does not work for time prediction of cases on which it was not trained, such as those with different disturbances. Rather, we seek an approach that can work over many different cases, including those on which it has not been trained. To subsume a wider set of cases, one could learn an approximate set of nonlinear governing equations for the state from the available training data, e.g., by using the sparse identification of nonlinear dynamics \cite{doi:10.1073/pnas.1517384113}, which approximates the governing equation by a set of candidate nonlinear functions and identifies the dominant terms by sparse regression, or by a neural ODE \cite{LINOT2023111838,linot2023turbulence,doi:10.1098/rspa.2022.0297}, which learns the underlying nonlinear functions automatically. To ensure generality for capturing possible dependencies on historical behavior and achieving robust long-time predictions, we use a recurrent neural network (RNN). specifically, we use long short-term memory (LSTM) architecture \cite{10.1162/neco.1997.9.8.1735}, which alleviates issues of gradient exploding/vanishing problem during training of other RNN versions \cite{annurev:/content/journals/10.1146/annurev-fluid-010719-060214}. The LSTM has recently been successful in predicting dynamics in many complex nonlinear environments \cite{doi:10.1098/rspa.2017.0844,VLACHAS2020191}.

Therefore, in the present study, we incorporate a novel model into the MBRL structure where the model consists of two parts: the PA-AE and the LDM. We train the RL agent directly from the model, but crucially, the evaluation of the trained agent is performed in the full CFD simulation. In Section \ref{sec:methodology}, we provide details of the model structure, including the approach of employing PA-AE to perform flow state abstraction by compressing the flow state into a three-dimensional latent space, and the utilization of LDM, which consists of a stacked LSTM, to learn the latent dynamics such that it can output time-series latent variable predictions given initial conditions\revision{,} any control input sequence\revision{, and, if applicable, auxiliary parameters}. Then, in Section \ref{sec:model_application}, the training and validation of the proposed models are performed in \revision{the scenario of} a pitching airfoil in a highly disturbed environment. Finally, in Section \ref{sec:MBRL_realization}, we \revision{implement} the MBRL \revision{to mitigate the lift variation through pitching control} by incorporating the models trained in Section \ref{sec:model_application} into the RL environment. 

\section{Methodology}\label{sec:methodology}
 In this section, our goal is to develop a low-order model of the dynamics of an aerodynamic flow that can be used as a surrogate flow environment for the purpose of training a reinforcement learning policy. As such, this model must be capable of advancing the state of the environment (i.e., our latent-space representation of the flow) \revision{and} describing the flow's response to actions, such as some control input applied to a wing. First, we will introduce a novel LDM that is based on the PA-AE to learn the flow dynamics. Subsequently, we will utilize the Twin Delayed Deep Deterministic policy gradient (TD3) \revision{\cite{DBLP:journals/corr/abs-1802-09477}} deep reinforcement learning algorithm to approximate the optimal policy from the model environment.

\subsection{Model Overview}\label{subsection:Model_Overview}
\begin{figure}[hbt!]
\centering
\includegraphics[width=1.0\textwidth]{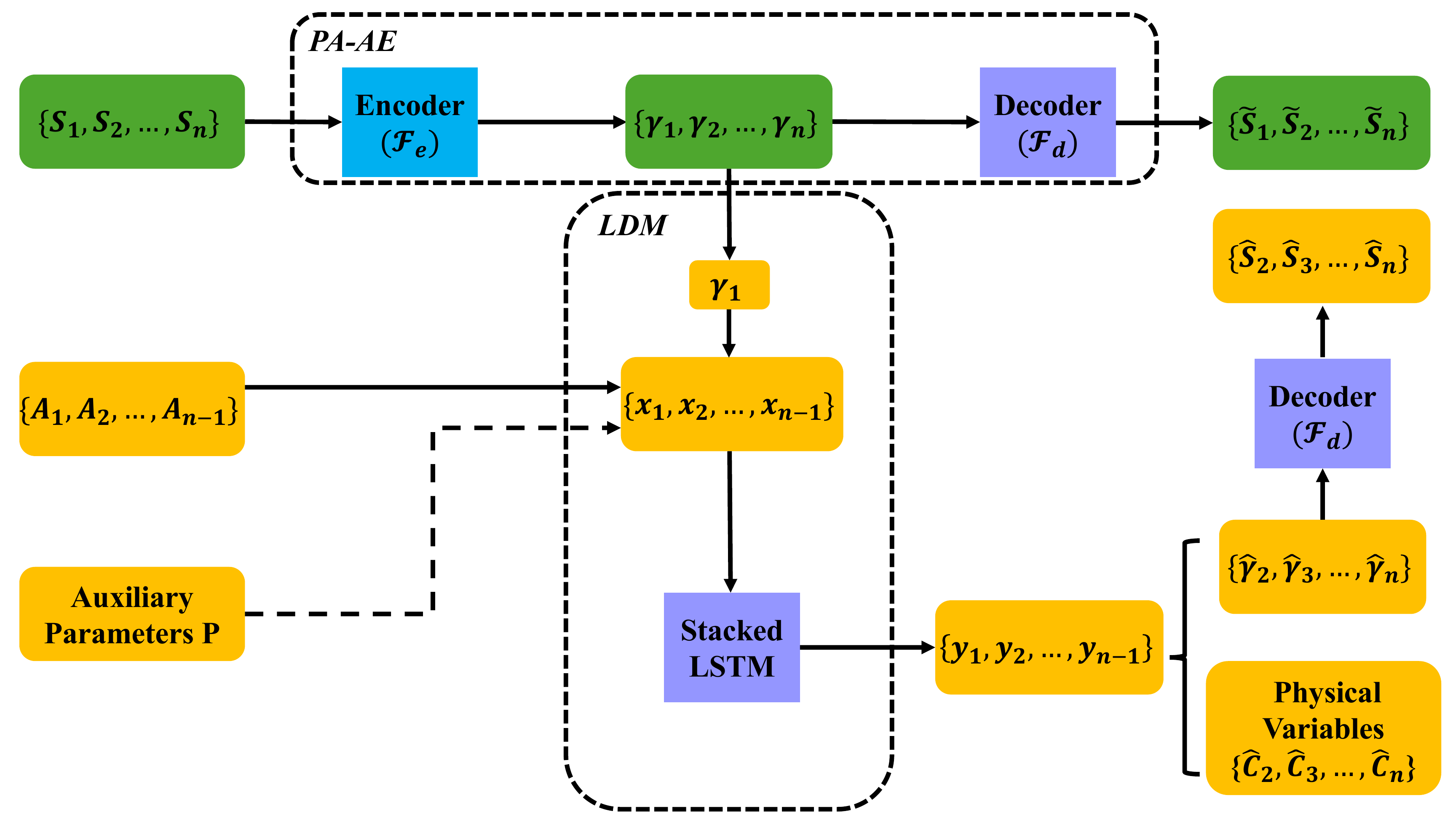}
\caption{\revision{Overview of the model structure (details of the stacked LSTM are provided in Fig. \ref{LSTM})}}
\label{Model_Overview}
\end{figure}

The model we propose in this work integrates a PA-AE and a LDM, as illustrated in \fig~\ref{Model_Overview}. The sequence $\{\mathbf{S}_1, \mathbf{S}_2, \dots, \mathbf{S}_n\}$ represents a time sequence of full flow states for a simulation case, generated from a corresponding sequence of actions, $\{A_1, A_2, \dots, A_n\}$, applied to the system. Each state comprises a CFD field variable of an incompressible flow, such as the fluid vorticity or, alternatively, velocity and pressure; for the scope of this work, we focus exclusively on the vorticity field. The PA-AE compresses each high-dimensional flow state $\mathbf{S}_k$ (of order $\mathcal{O}(10^5)$ in a typical 2D simulation) into a very low-dimensional latent variable $\boldsymbol{\gamma}_k$ using its encoder component. The decoder component subsequently reconstructs the flow state from this latent variable, yielding $\mathbf{\tilde{S}}_k$; a companion network augments this by predicting another important physical variable, such as lift or power coefficient. We thereby obtain a training set composed of comprehensive histories of latent variables in the form of sequences $\{\boldsymbol{\gamma}_1, \boldsymbol{\gamma}_2, \dots, \boldsymbol{\gamma}_n\}$, with each sequence uniquely representing a different simulation case generated from a corresponding action sequence, $\{A_1, A_2, \dots, A_{n-1}\}$. Additionally, we may also include auxiliary parameters $\mathbf{P}$, such as those describing an environmental disturbance (e.g., a gust). The inclusion of $\mathbf{P}$ depends on whether an initial flow state and an action sequence suffice to distinguish the corresponding simulation case from others.

Upon incorporating the initial latent variable $\gamma_1$, the corresponding action sequence $\{A_1, A_2, \dots, A_{n-1}\}$, and, potentially, the auxiliary parameters $\mathbf{P}$, the input sequence $\{\mathbf{x}_1, \mathbf{x}_2, \dots, \mathbf{x}_{n-1}\}$ is systematically arranged and fed into a stacked LSTM network. This process yields an output sequence $\{\mathbf{y}_1, \mathbf{y}_2, \dots, \mathbf{y}_{n-1}\}$, which includes both the predicted sequence of latent variables $\{\boldsymbol{\hat{\gamma}}_2, \boldsymbol{\hat{\gamma}}_3, \dots, \boldsymbol{\hat{\gamma}}_n\}$ for subsequent time steps and a sequence of predicted physical variables, such as the lift coefficient $C_l$ or power coefficient $C_p$. For validation purposes, the decoder part of the PA-AE is then used to reconstruct the flow state from the predicted latent variable sequence. These predicted flow states are then compared with the true flow states obtained from flow simulations to assess model accuracy. Additionally, the sequences of predicted physical variables can also be compared with the true physical variable sequences. After validation, we can apply this model for training the MBRL policy. Each of these aspects will be discussed in the remainder of this section.
\subsection{Physics-Augmented Autoencoder Framework}\label{sec:PA-AE}
\begin{figure}[hbt!]
\centering
\includegraphics[width=1.0\textwidth]{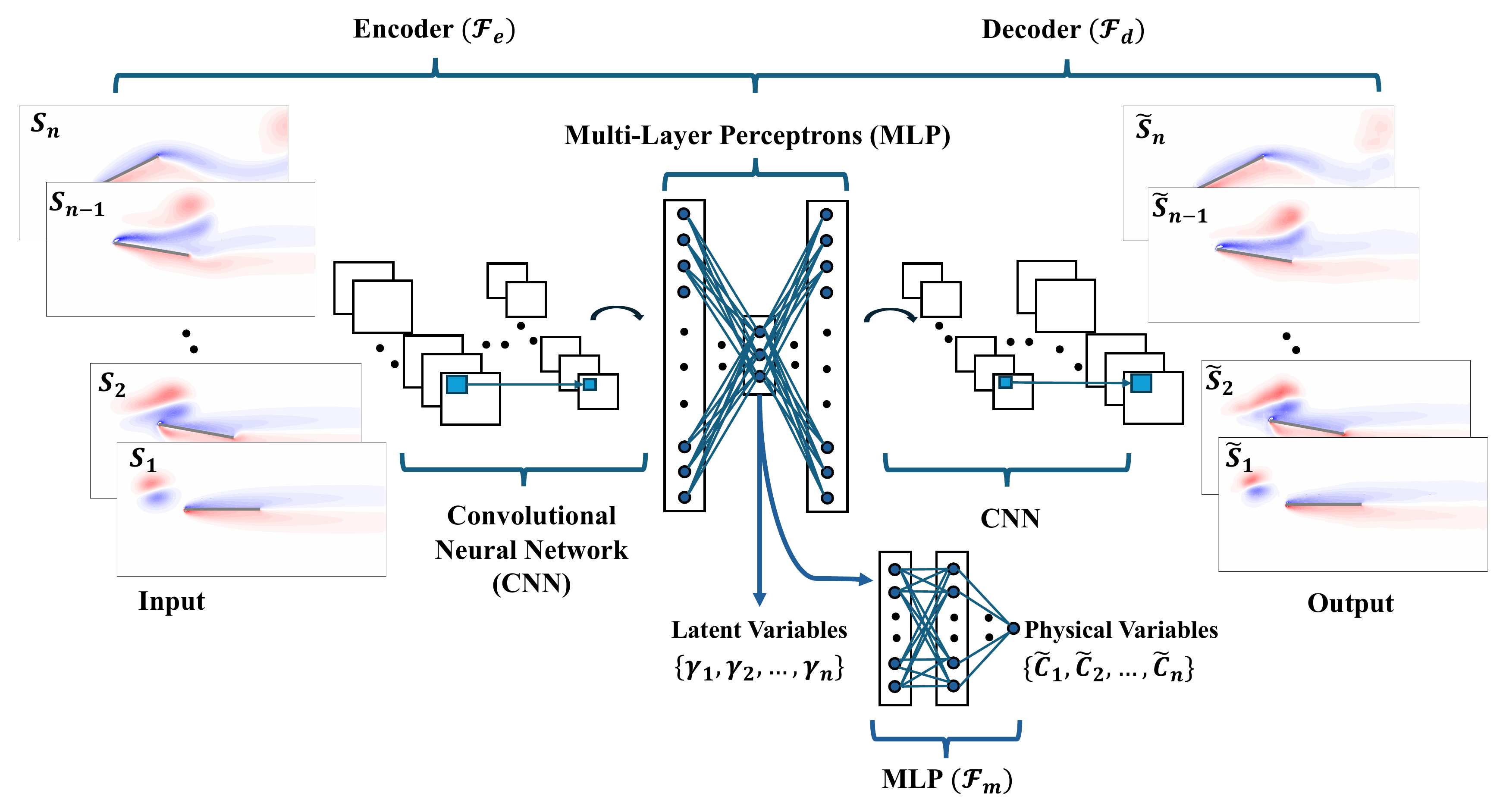}
\caption{The structure of the PA-AE, adapted from Fukami and Taira \cite{Fukami2023}}
\label{Auto_Encoder}
\end{figure}

Inspired by the work of Fukami and Taira \cite{Fukami2023}, who demonstrated that the fundamental physics underlying extreme aerodynamics of a two-dimensional flow past a wing could be effectively represented within a three-dimensional manifold using a lift-augmented autoencoder, we adapt this concept to develop our PA-AE, as illustrated in \fig~\ref{Auto_Encoder}. It is important to note that, while our PA-AE extends the utility to other physical variables beyond lift, this study does not compare its performance directly with that of traditional autoencoders.

In our approach, the PA-AE $\mathcal{F}$ employs a convolutional neural network (CNN) to process the input vorticity field $\mathbf{S}_k$. After it is fed into the CNN layers, the data is flattened and subsequently passed through a multi-layer perceptron (MLP), which compresses it into a three-dimensional latent variable $\boldsymbol{\gamma}_k$. This compression is executed by the encoder component, $\mathcal{F}_e$, of the PA-AE. The decoder component, $\mathcal{F}_d$, then begins by expanding the latent variable into a higher-dimensional space using an MLP; the result is reshaped and further processed by another CNN to reconstruct the vorticity field $\mathbf{\tilde{S}}_k$. Simultaneously, the latent variable is processed by another MLP to reconstruct \revision{a} designated physical variable $C_k$. Therefore, this PA-AE is seeking to achieve
\begin{equation}\label{eqn:Autoencoder_target}
    (\mathbf{S}_k,C_k)\approx (\tilde{\mathbf{S}}_k,\tilde{C}_k) = \mathcal{F}(\mathbf{S}_k;\mathbf{W})= (\mathcal{F}_d(\boldsymbol{\gamma}_k),\mathcal{F}_m(\boldsymbol{\gamma}_k))=(\mathcal{F}_d(\mathcal{F}_e(\mathbf{S_k})),\mathcal{F}_m(\mathcal{F}_e(\mathbf{S_k})))\revision{,}
\end{equation}
In particular, the PA-AE is trained by finding the optimal weights $\mathbf{W}$ (via the Adam optimizer) that minimize the loss function $\mathcal{J}$:
\begin{equation}\label{eqn:Autoencoder_loss}
    \mathbf{W}=\underset{\mathbf{W}}{\text{argmin}}(\mathcal{J})=\underset{\mathbf{W}}{\text{argmin}}(||\mathbf{S_k}-\mathbf{\tilde{S}}_k||_2+\beta||C_k-\tilde{C}_k||_2).
\end{equation}
In this study, the PA-AE trained with $\beta=0.05$ is used for further LDM and RL training, though other values have been tried but show inferior performance. Since most \revision{unsteady} aerodynamic flows exhibit rich nonlinear behavior, we choose to use a hyperbolic tangent activation function $f(z)=(e^{z}-e^{-z})/(e^{z}+e^{-z})$, \revision{which has been shown to work well for capturing flow features when the flow disturbances are strong \cite{Fukami2023}}. As a result, the latent space components are each restricted to the range $[-1,1]$. The details about the specific structure of the PA-AE is shown in the Appendix.

\subsection{Latent Dynamic Model Framework}\label{sec:LDM}
After learning the low dimensional representation of the flow state, we then proceed to the dynamics learning. In this subsection, we describe the LDM that leverages the capabilities of a stacked LSTM network to learn the dynamics in the latent space, as depicted in \fig~\ref{LSTM}. The LSTM is a powerful neural network model that is capable of learning order dependence in sequence prediction problems, making it ideal for processing and learning from time-series data.

As discussed in Section \ref{subsection:Model_Overview}, the composition of the input sequence \( \mathbf{x}_t \) is contingent upon the specifics of the problem setting. We need the LDM to be capable of predicting both the low-dimensional representations of all subsequent flow states \( \{\boldsymbol{\hat{\gamma}}_2, \boldsymbol{\hat{\gamma}}_3, \ldots, \boldsymbol{\hat{\gamma}}_n\} \) and the corresponding physical variables for later time steps \( \{\hat{C}_2, \hat{C}_3, \ldots, \hat{C}_n\} \), given an input sequence. Therefore, the formulation of the input sequence is as follows:
\begin{equation}\label{eqn:LDM_input}
  \{\mathbf{x}_1, \mathbf{x}_2, \ldots, \mathbf{x}_{n-1}\} = 
  \begin{cases} 
    \{(\boldsymbol{\gamma}_1, A_1), (\boldsymbol{\gamma}_1, A_2), \ldots, (\boldsymbol{\gamma}_1, A_{n-1})\} & \text{if condition 1} \\
    \{(\boldsymbol{\gamma}_1, \mathbf{P}, A_1), (\boldsymbol{\gamma}_1, \mathbf{P}, A_2), \ldots, (\boldsymbol{\gamma}_1, \mathbf{P}, A_{n-1})\} & \text{if condition 2}
  \end{cases}
\end{equation}
where:
\begin{itemize}
    \item Condition 1 stipulates that the initial flow representation \( \boldsymbol{\gamma}_1 \) combined with the action sequence suffices to predict the flow state at all subsequent time steps.
    \item Condition 2 asserts that the inclusion of auxiliary parameters \( \mathbf{P} \) is essential for a comprehensive description of the flow dynamics. \revision{The auxiliary parameters \( \mathbf{P} \) are included for all the gust cases in this current work.}
\end{itemize}
The output sequence is then defined as:
\begin{equation}\label{eqn:LDM_output}
\{\mathbf{y}_1, \mathbf{y}_2, \ldots, \mathbf{y}_{n-1}\} = \{(\boldsymbol{\hat{\gamma}}_2, \hat{C}_2), (\boldsymbol{\hat{\gamma}}_3, \hat{C}_3), \ldots, (\boldsymbol{\hat{\gamma}}_n, \hat{C}_n)\}
\end{equation}
This structured output not only ensures the model's ability to predict future states but also to quantify the designated physical variable trajectory.

This LSTM-based LDM \revision{enables us to make accurate predictions over long time horizons and} also allows for the incorporation of other physical variables into the machine learning model to facilitate additional \revision{distinguishability of trajectories in} the model's prediction. For validation purposes, the model's efficacy is not limited to comparisons between predicted and actual values of latent and physical variables: the predicted latent variables generated by the LDM can also be decoded using the PA-AE's decoder component to reconstruct the instantaneous full flow state. These reconstructed states are then compared with the actual flow states obtained from the original CFD simulations, providing a comprehensive evaluation of the model’s predictive accuracy. 

\subsection{Reinforcement Learning Framework}\label{sec:RL framework}
\begin{figure}[hbt!]
\centering
\includegraphics[width=1.0\textwidth]{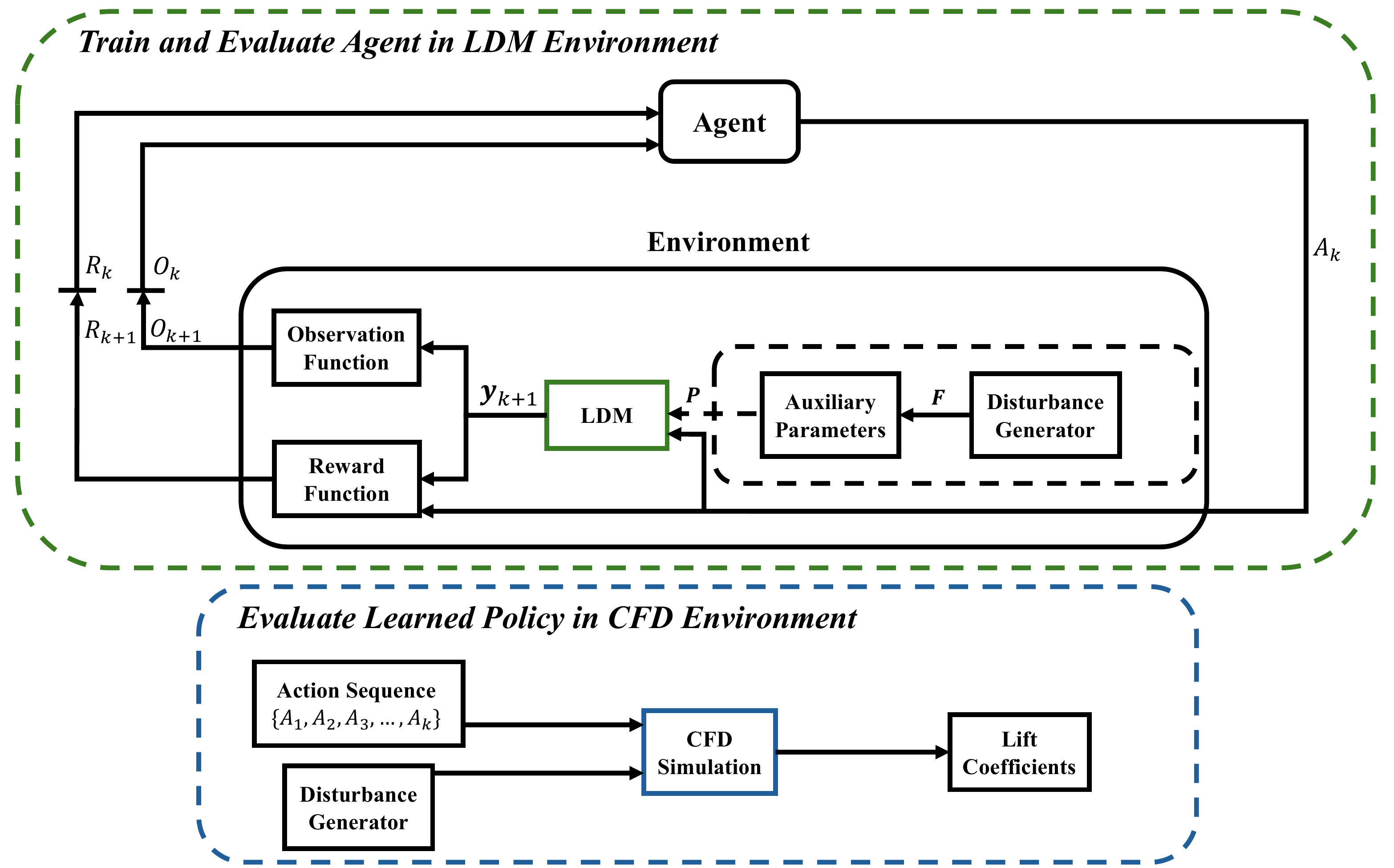}
\caption{\revision{RL Overview, adapted from Beckers and Eldredge \cite{beckers2024deepreinforcementlearningairfoil}}}
\label{RL_Overview}
\end{figure}

After establishing the LDM, we encapsulate it within the RL framework, as illustrated in \fig~\ref{RL_Overview}. The framework consists of two primary components: the Agent and the Environment. Here, the Agent functions as the controller, while the Environment incorporates the LDM and the observation and reward functions. Notably, the Environment may include a disturbance generator, which varies depending on the specific flow problem addressed. Conventionally, model-free RL that utilizes CFD simulations predicts the next state \(\mathbf{S}_{k+1}\) from the current state \(\mathbf{S}_k\) and the action \(A_k\). Similarly, a typical low-dimensional model updates to \(\boldsymbol{\hat{\gamma}}_{k+1}\) from \(\boldsymbol{\hat{\gamma}}_k\) and \(A_k\). However, as outlined in Section \ref{sec:LDM}, our LDM advances the state to \(\boldsymbol{\hat{\gamma}_{k+1}}\), \revision{given} the initial condition \(\boldsymbol{\gamma}_1\)\revision{,} the action history up to that point, i.e., \(\{A_1, A_2, \ldots, A_k\}\), and auxiliary parameters \(\mathbf{P}\) when necessary. The latest observation \(O_{k+1}\) and reward \(R_{k+1}\) are determined from the updated state \(\boldsymbol{\hat{\gamma}}_{k+1}\) and physical variable \(\hat{C}_{k+1}\), via the observation and reward functions, respectively. These functions are chosen specifically for a problem and will be further detailed in the example in Section \ref{sec:MBRL_realization}. The agent chooses an action from the current policy, given the current observation. Due to the complexity of flow environments, such as the presence of random upstream gusts, off-policy algorithms are particularly valuable. Their ability to learn from experiences generated under any policy, which may encompass a wide range of environmental conditions, significantly enhances both robustness and efficiency. We follow Beckers and Eldredge \cite{beckers2024deepreinforcementlearningairfoil} and use the TD3 algorithm \cite{DBLP:journals/corr/abs-1802-09477}, which utilizes an off-policy, continuous action space framework.

\revision{Once the RL agent is trained in the LDM environment, we evaluate the learned policy in both the LDM environment and the CFD environment. Specifically, we first evaluate the learned policy in the LDM environment under any gust, and collect the resulting action sequence $\{ A_1, A_2, \ldots, A_k\}$ and the corresponding gust parameters. Then, this same action sequence is applied to the CFD simulation under the corresponding gust when we do the evaluation in that environment.}

\section{Model Application}\label{sec:model_application}
In this section, we apply the PA-AE and LDM framework detailed in Sections~\ref{sec:PA-AE} and \ref{sec:LDM} to \revision{a representative scenario:} a pitching airfoil in a highly disturbed environment. \revision{This application aims} to demonstrate the accuracy and robustness of our modeling framework \revision{in gust environments}. \revision{We also study the application of our modeling framework to a vertical-axis wind turbine (VAWT) in a disturbance-free environment to show the generalizability of our framework. Specifically, we demonstrate that the autoencoder augmentation can be applied not only to the lift coefficient but also to the power coefficient, and that auxiliary parameters are unnecessary in the absence of gusts in the flow environment. However, to maintain clarity and conciseness in the main text, we present this second example in the Appendix and refer interested readers there for details.}

\subsection{Problem Set-Up and Simulation}\label{sec:pitch_distb_set_up}
\begin{figure}[hbt!]
\centering
\includegraphics[width=1.0\textwidth]{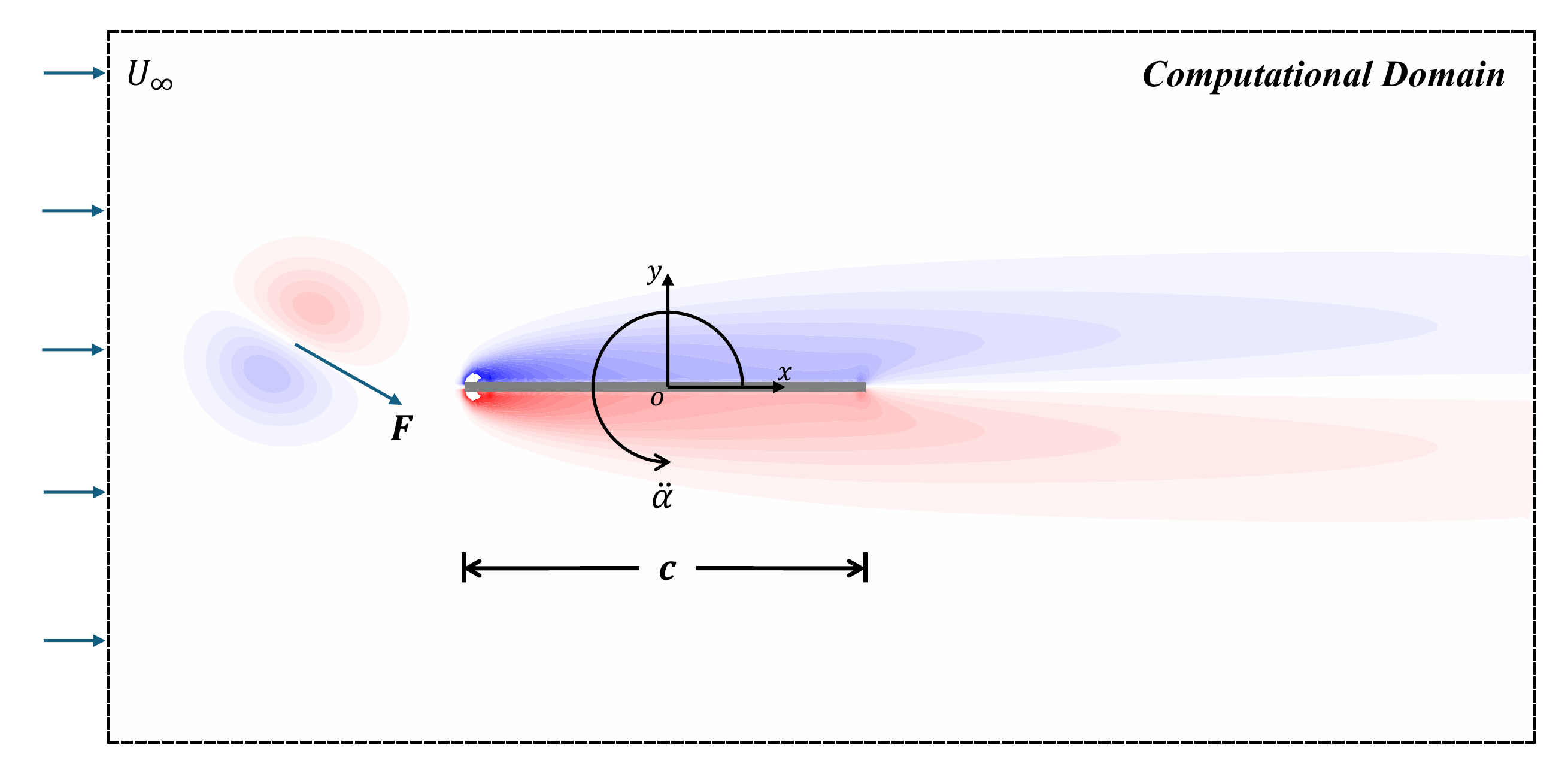}
\caption{The schematic figure of the problem set-up for the pitching airfoil in the gust environment in the form of spatiotemporal Gaussian force applied to the fluid.}
\label{pitch_distb_schematic}
\end{figure}

In this context, we consider a flat plate airfoil of chord length $c$, at an initial angle $\alpha=0^\circ$ with the $x$ axis (positive in the counter-clockwise direction), in a fluid of kinematic viscosity $\nu$ in a constant uniform flow $U_{\infty}$ in the $x$ direction. The airfoil can undergo prescribed pitching about a pivot point located in the center of the chord, as depicted in \fig~\ref{pitch_distb_schematic}. We introduce convecting gusts in the form of vortices generated from a Gaussian forcing field $\mathbf{F}$ \revision{which acts as a body force term in the governing equation}, with its spatial center located upstream of the airfoil, described by 
\begin{equation}\label{eqn:Gaussian}
    \mathbf{F}(x,y,t) = \rho U_{\infty}^2 c \frac{(D_x,D_y)}{\pi^{3/2}\sigma_x \sigma_y \sigma_t} \exp\left[-\frac{(x-x_0)^2}{\sigma_x^2}-\frac{(y-y_0)^2}{\sigma_y^2}-\frac{(t-t_0)^2}{\sigma_t^2}\right],
\end{equation}
where we fix $D_x=1.2$, $x_0=-1.0 c$, $\sigma_t=0.2 c/U_\infty$, and $t_0=0.5 c/U_\infty$. Also, we restrict $\sigma=\sigma_x=\sigma_y$ in this study such that there are only three varying gust parameters, $D_y$, $\sigma$, and $y_0$. Specifically, $D_y$, $\sigma$, and $y_0$ are uniformly sampled from $[-1,1]$, $[0.05c,0.1c]$, and $[-0.25c,0.25c]$, respectively. \revision{A Gaussian forcing field acts as a source term in the Navier--Stokes equations, and thus, generates a distributed region of vorticity in the fluid. The Gaussian spatial and temporal distributions ensure that this new vorticity comes in the form of vortices that are localized (but continuous) in space and time. The generated vorticity is subject to the Navier--Stokes equations, like any other flow feature, e.g., it convects under the flow. It is worth noting that the reason why we introduce the Gaussian forcing field as a function of time is that we seek to continuously introduce these gusts rather than create them impulsively, e.g., through initial conditions. It should be noted that, by introducing gusts in this manner, the governing equations are different for each case with different Gaussian forcing field parameters.} The pitching, which will constitute the action for the later control effort, is commanded via prescribed angular acceleration $\ddot{\alpha}$ around the pivot point. Though we will eventually seek a control policy that minimizes the lift coefficient variations about a prescribed lift reference, shown in Section~\ref{sec:MBRL_realization}, we will first use a random policy here to generate the model training dataset, comprising the vorticity field $\omega_k$, lift coefficient $C_{L_k}$, and the action $A_k$, at chosen time steps, as well as the corresponding auxiliary parameter $\mathbf{P}=(D_y,\sigma,y_0)$ for each simulation case.

We use a viscous incompressible flow solver based on an immersed boundary projection method \cite{eldredge2022method} to perform the simulation at Reynolds number $Re=U_\infty c/\nu=200$, as shown in \fig~\ref{pitch_distb_schematic}. From hereon, we will report time in convecting units, $t^* = tU_\infty/c$, and spatial coordinates in chord lengths, $x^* = x/c$ and $y^* = y/c$. Each simulation case lasts for at most 3 convective time units, starting from a fully developed flow about a flat plate at $\alpha=0^\circ$. We set up the pivot point as the origin of the computational domain, which spans $[-1.395,2.175]$ in the $x$ direction and $[-0.885,0.885]$ in the $y$ direction and employs free-space boundary conditions on all boundaries. The flow solver employs a vorticity-based approach and utilizes a lattice Green’s function to solve the Poisson equation, allowing for a more compact domain compared to traditional computational aerodynamics studies. In this computational domain set up, the vortex gust and the resulting airfoil response are ensured to remain between the upper and lower boundary and freely convect through the downstream boundary. We employ a uniform Cartesian grid with grid size $\Delta x^*=0.015$, corresponding to a grid of size $240\times 120$, and a time step size $\Delta t^* = 0.006$. In order to simulate a more practical control problem with limited bandwidth, we restrict the pitching action $\ddot{\alpha}$ to be piecewise constant over 0.03 convective time units, which means that the $\ddot{\alpha}$ would be resampled randomly each $5\Delta t^*$ time units, and there are at most 100 action steps for each simulation. We uniformly save the data every $0.03$ convective time units, so there are $101$ vorticity snapshots, $101$ lift coefficient data points, and $100$ action points saved at most for each simulation case.

The angular acceleration of the plate is drawn from a random policy
\begin{equation}\label{eqn:random_policy}
  \ddot{\alpha}_k = 
  \begin{cases} 
    0.1\times \text{randn()} & \text{if $1\leq k < 11$}, \\
    5\times \text{randn()} & \text{if $11\leq k < 51$}, \\
    \text{randn()} & \text{if $51\leq k\leq 100$},
  \end{cases}
\end{equation}
where the randn() generates a normally-distributed random number with mean 0 and standard deviation 1, and $k$ represents the action time step. \revision{We chose this policy to help maintain a small magnitude of angular acceleration during the early times (prior to 0.3 convective times) when the gust has not yet reached the airfoil, and then increase it significantly during the peak gust interaction (between 0.3 and 1.5 convective time units), and finally, reduce it afterward when the gust has convected past the airfoil.} To avoid unrealistic configurations, we terminate the simulation early if $|\alpha| > 45^\circ$. \revision{The policy above helps reduce the chance for early truncation, allowing more full-length cases to be collected into the training set.} As a result, we collect a total of $1013$ simulation cases (with a total of $72900$ vorticity snapshots) for the training and validation purpose in this study, of which 650 are truncated early.

\begin{figure}[hbt!]
\centering
\includegraphics[width=1.0\textwidth]{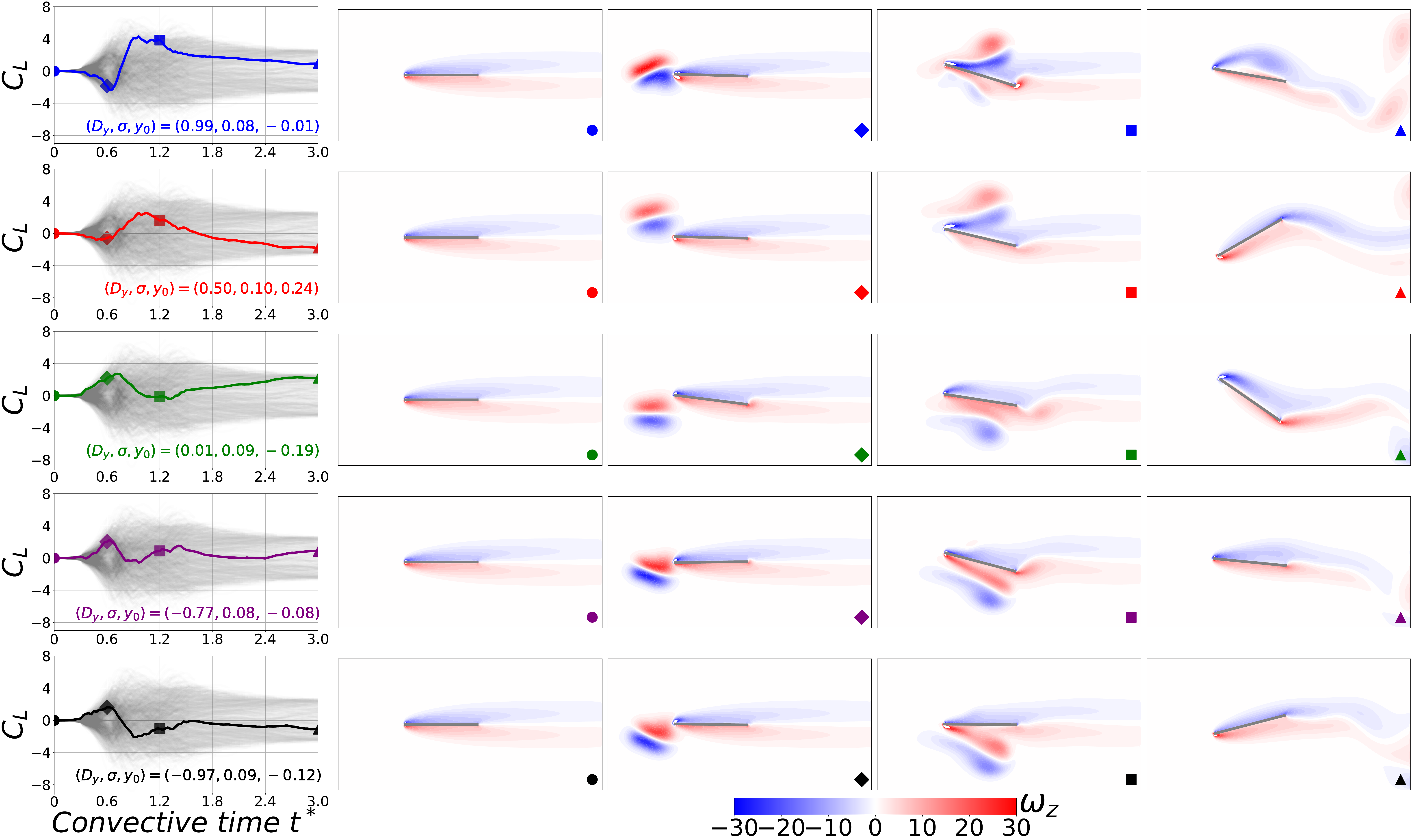}
\caption{Examples of the lift response and the vorticity field. The lift response for all cases present in the study is shown in \revision{light grey color}. The snapshots of the vorticity field correspond to the times indicated by the symbols $\circ$, $\Diamond$, $\square$, $\triangle$, filled with different colors representing different cases.}
\label{pitch_distb_original}
\end{figure}

Since the gust parameters and the flat plate pitching actions vary from case to case, the flow fields and the corresponding lift coefficients also vary significantly for different cases, as depicted in \fig~\ref{pitch_distb_original}, due to the nonlinear vortex dynamics. When the gust passes the pitching airfoil, it may induce new vortex structures depending on how the flat plate encounters the incoming gust. For some of these cases---for example, when a gust dipole is cleaved in two by the airfoil, as in the first case in \fig~\ref{pitch_distb_original}---the modeling and control is expected to be particularly challenging. It is important to mention that the initial vorticity field is \revision{the same for all collected cases nearly to within machine error,} in spite of the different gust parameters, since the gust forcing in \revision{Eq.}~\eqref{eqn:Gaussian} does not start significantly until $t^* = 0.3$. This is analogous to a more realistic gust encounter, in which the airfoil would be in quiescent surroundings prior to the encounter. As a result, \revision{the PA-AE compresses the initial fields of all different cases into the same latent variable,} as we will show in Sec. \ref{sec:pitch_distb_PA-AE}. That is, the initial state is insufficient to distinguish the disturbed environments of different cases, creating the need to augment this state with the gust parameter $\mathbf{P}$ for LDM training purposes.

\subsection{PA-AE Training and Validation}\label{sec:pitch_distb_PA-AE}
After collecting the $1013$ simulation cases, we use the obtained $72900$ vorticity snapshots and the corresponding lift coefficient histories of these cases to train the PA-AE to compress the vorticity data, each with a dimension of $240\times120$, to a latent space with a dimension of three. In order to obtain a model with good generalization, we randomly split the dataset into a training dataset comprising $70\%$ and a validation dataset of $30\%$. The loss function is defined as in \revision{Eq.}~\eqref{eqn:Autoencoder_loss}, since Fukami and Taira \cite{Fukami2023} have been successful in reconstructing the vorticity field via a similar autoencoder with the same loss function.

\begin{figure}[hbt!]
\centering
\includegraphics[width=0.5\textwidth]{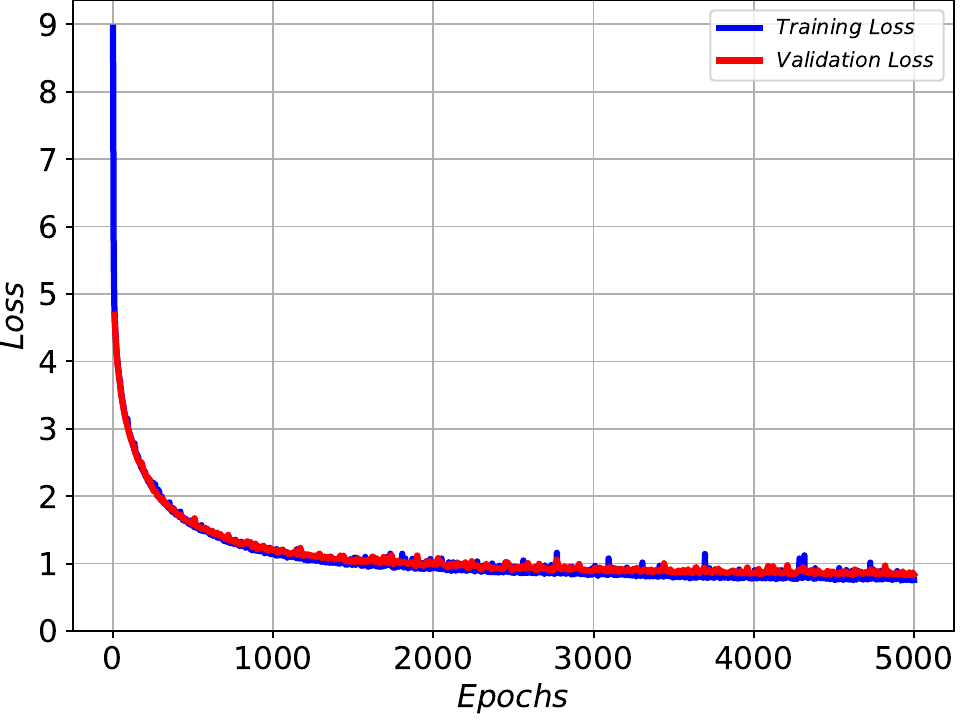}
\caption{The training loss and validation loss history for the PA-AE for the problem of pitching airfoil with gust environment.}
\label{pitch_distb_AE_loss}
\end{figure}

The training loss and validation loss are shown in \fig~\ref{pitch_distb_AE_loss}. The validation loss is calculated every $10$ epochs in order to save computational time. \revision{In machine learning, each epoch corresponds to one complete iteration over the training dataset during model optimization}. We can observe that both the training loss and the validation loss decrease as the epoch increases, which shows that the model is not overfitting and will have good generalization behavior because the validation dataset consists of all the unseen vorticity snapshots. Based on this loss profile, we use the model generated at the $5000$th epoch to do further study. For reference, the PA-AE training in this case takes approximately $50$ hours on an NVIDIA A100 GPU card.

\begin{figure}[hbt!]
\centering
\includegraphics[width=1.0\textwidth]{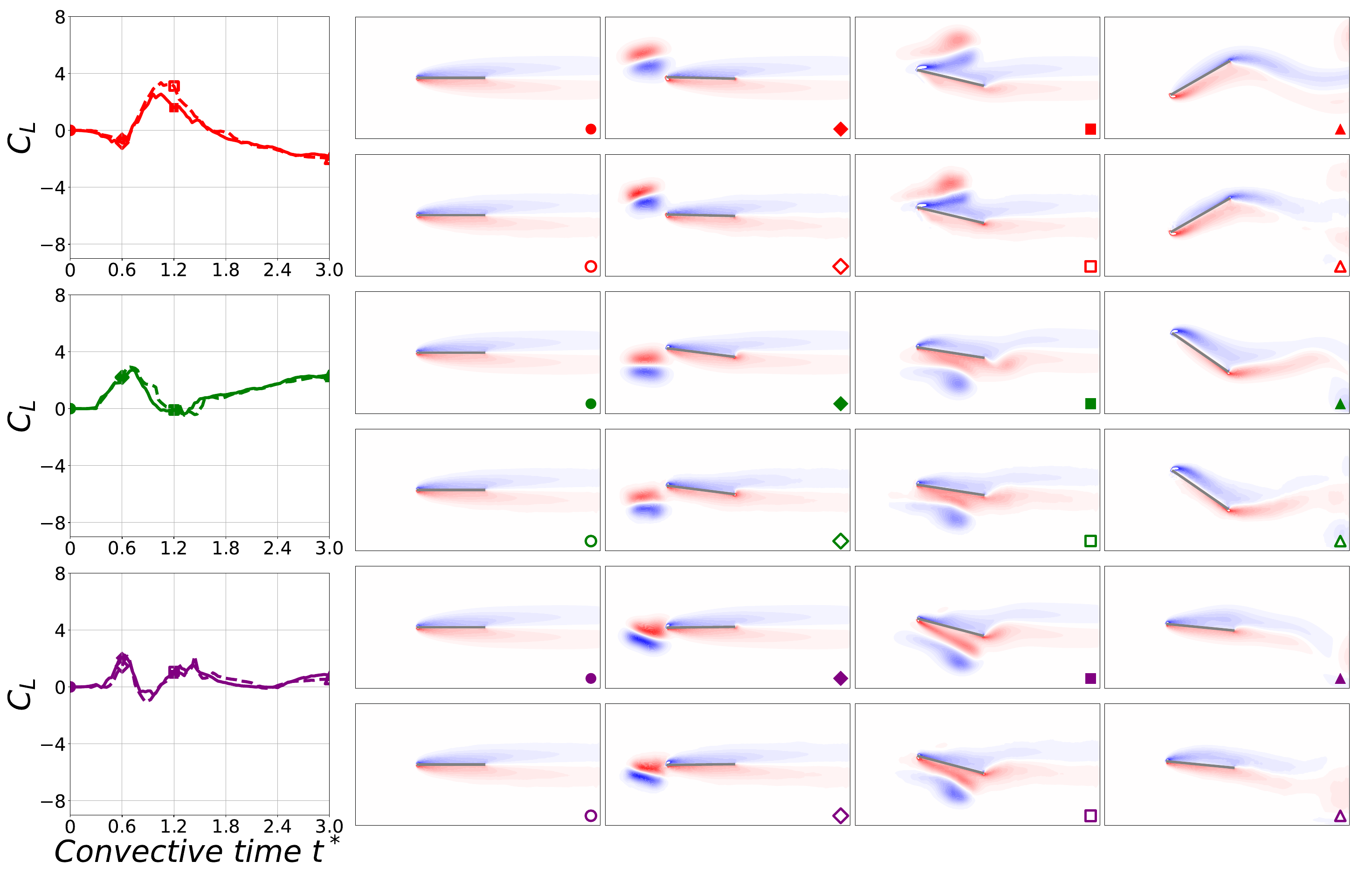}
\caption{Three examples of the PA-AE reconstruction of the lift coefficient and corresponding vorticity field. The solid line \protect\tikz[baseline]{\protect\draw[solid, thick] (0,.8ex) -- (0.3,.8ex);} represents the ground truth lift, and the dashed line \protect\tikz[baseline]{\protect\draw[dashed, thick] (0,.5ex) -- (0.3,.5ex);} represents the reconstructed lift from PA-AE. The truth vorticity field and the reconstructed vorticity field are labeled by solid and blank symbol, respectively. The convention of colors and symbols used here is consistent with the one in \fig~\ref{pitch_distb_original}.}
\label{pitch_distb_AE_reconstruction}
\end{figure}

Since it is difficult to assess the performance of the reconstruction from only the loss profiles, representative examples of the vorticity field and lift coefficient reconstruction are given in \fig~\ref{pitch_distb_AE_reconstruction}. We observe that the PA-AE not only reconstructs the gust structure very well, but also has the ability to capture the angle of attack of the flat plate by identifying the boundary layer structures. However, the wake structure for convective time close to $3$ exhibits some flaws in the reconstruction. This is reasonable since we have indicated in Section~ \ref{sec:pitch_distb_set_up} that there is a relative dearth of training data for large times because of the potential truncation of the simulation. Nevertheless, even with less data, the PA-AE is still capable of reconstructing the boundary layer structures and the correct angle of attack of the flat plate for these large $t^*$ snapshots. Furthermore, the PA-AE also shows good performance in the reconstruction of the lift coefficients for most of time, though it exhibits relatively large bias error for intervals with strong gust-plate interactions. The reason may be that the lift coefficient is sensitive to the structure and position of the gust during a gust-plate interaction, so that a slight error in reconstructing the structure and position of the gust may result in a visually larger error in lift coefficient reconstruction. This bias could be potentially reduced by increasing the size of the training dataset at the cost of larger computational time for obtaining data and training the PA-AE model, or by fine-tuning the hyper-parameters of the PA-AE model. We stress that the current hyper-parameters are not necessarily optimal.

\begin{figure}[hbt!]
\centering
\includegraphics[width=0.65\textwidth]{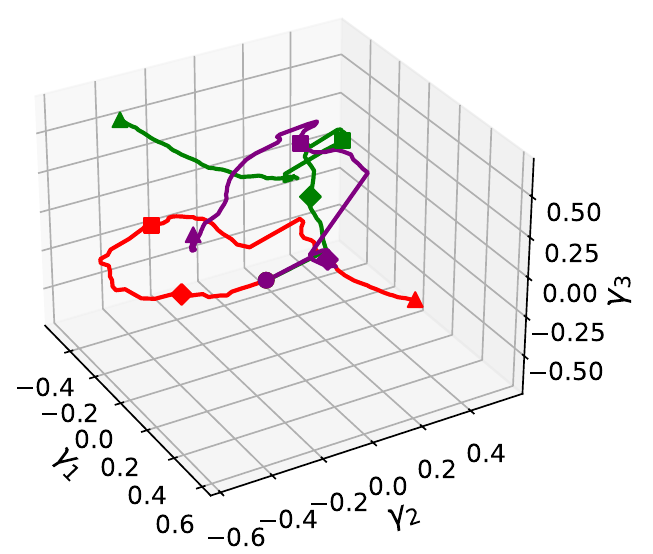}
\caption{Visualization of the latent variables $\boldsymbol{\gamma}_k=(\gamma_{1_k},\gamma_{2_k},\gamma_{3_k})$ histories for the same three cases as shown in \fig~\ref{pitch_distb_AE_reconstruction}. The convention of colors and symbols indicated here is consistent with the one in \fig~\ref{pitch_distb_original} and \fig~\ref{pitch_distb_AE_reconstruction}}.
\label{pitch_distb_AE_latent}
\end{figure}

It is also worthwhile to inspect the latent variables trajectories, depicted in \fig~\ref{pitch_distb_AE_latent}, since \revision{they} may contain some key information about the flow field, as suggested by Fukami and Taira \cite{Fukami2023}. We first need to clarify that there is no clear physical meaning for these latent variables, and it is not the objective of this study to reveal any such meaning in detail. However, we will make a few observations. First, all latent variables trajectories start from the same point (the circular marker), which emphasizes the earlier discussion that subtle differences in the initial vorticity field are not detected by the PA-AE. Once the disturbances become significant we can observe several features from these trajectories. For example, the $\gamma_2$ seems to correlate with the relative location of the gust with the flat plate since, among the vorticity snapshots in \fig~\ref{pitch_distb_AE_reconstruction}, those labeled by the red diamond and red square are the only two in which the gust structure has a higher vertical position than the flat plate, and correspondingly, these are the only two labeled points that lie at lower values of $\gamma_2$ than the initial point. We can also observe that $\gamma_1$ may be correlated with the pitching angle of the flat plate, since the pitching angle of the flat plate in the vorticity snapshots labeled by the red and green triangles in \fig~\ref{pitch_distb_AE_reconstruction} has two opposite and extreme values, and similarly, the corresponding points in the latent space lie at two opposite ends of the $\gamma_1$ axis, in contrast to the other points at less extreme angles. \revision{The interpretability of the latent space and the mapping of uncertainty from the flow environment to this space remain an open topic and are the subject of ongoing investigation.}

\subsection{LDM Training and Validation}\label{sec:pitch_distb_LDM}
After we reduce the vorticity data from a $240\times 120$ space to the three-dimensional latent space by the PA-AE, we employ the LDM, as described in Section~\ref{sec:LDM}, to learn the latent dynamics. The ostensible objective is to predict the evolution of the flow field given the initial flow field and the action sequence. However, as discussed in Section~ \ref{sec:pitch_distb_PA-AE}, the initial flow field (or its corresponding latent space point) does not distinguish cases with different subsequent disturbances. Thus, it would be impossible to train the LDM to predict the $\boldsymbol{\gamma}_k$ trajectory accurately with only \revision{initial conditions and action sequences}. \revision{Specifically, we could imagine there are two different cases where each has a different Gaussian forcing field as the disturbance but has the exact same action sequence. Since the initial condition is the same, if we only provide the initial condition and the action sequence as the input to the LDM, it would output the same latent trajectory (state prediction sequence) for the two different cases, which is not what we expect. In order to make the LDM output latent trajectories that correspond to different gust scenarios, we need to also provide some auxiliary parameters to help the LDM distinguish each case. Since the differences between the two cases come from the different Gaussian forcing fields in the governing equations, we simply use the parameters of the Gaussian forcing field as the auxiliary parameters. This is why} we choose to set $\mathbf{P}=(D_y,\sigma,y_o)$ \revision{in this study}. Alternatively, if the disturbance\revision{s} were introduced in the form of an initial condition \revision{such as adding an initial vortex in the upstream}, or none were introduced at all, then the governing equations would remain invariant for all cases, \revision{then the initial condition} $\boldsymbol{\gamma}_1$ and \revision{action sequence }$\{A_1,A_2,...,A_k\}$ would be sufficient input for the LDM to learn the dynamics accurately. \revision{The details about the demonstration of a case with no disturbance are provided in the VAWT example in Appendix.}

\begin{figure}[hbt!]
\centering
\includegraphics[width=0.8\textwidth]{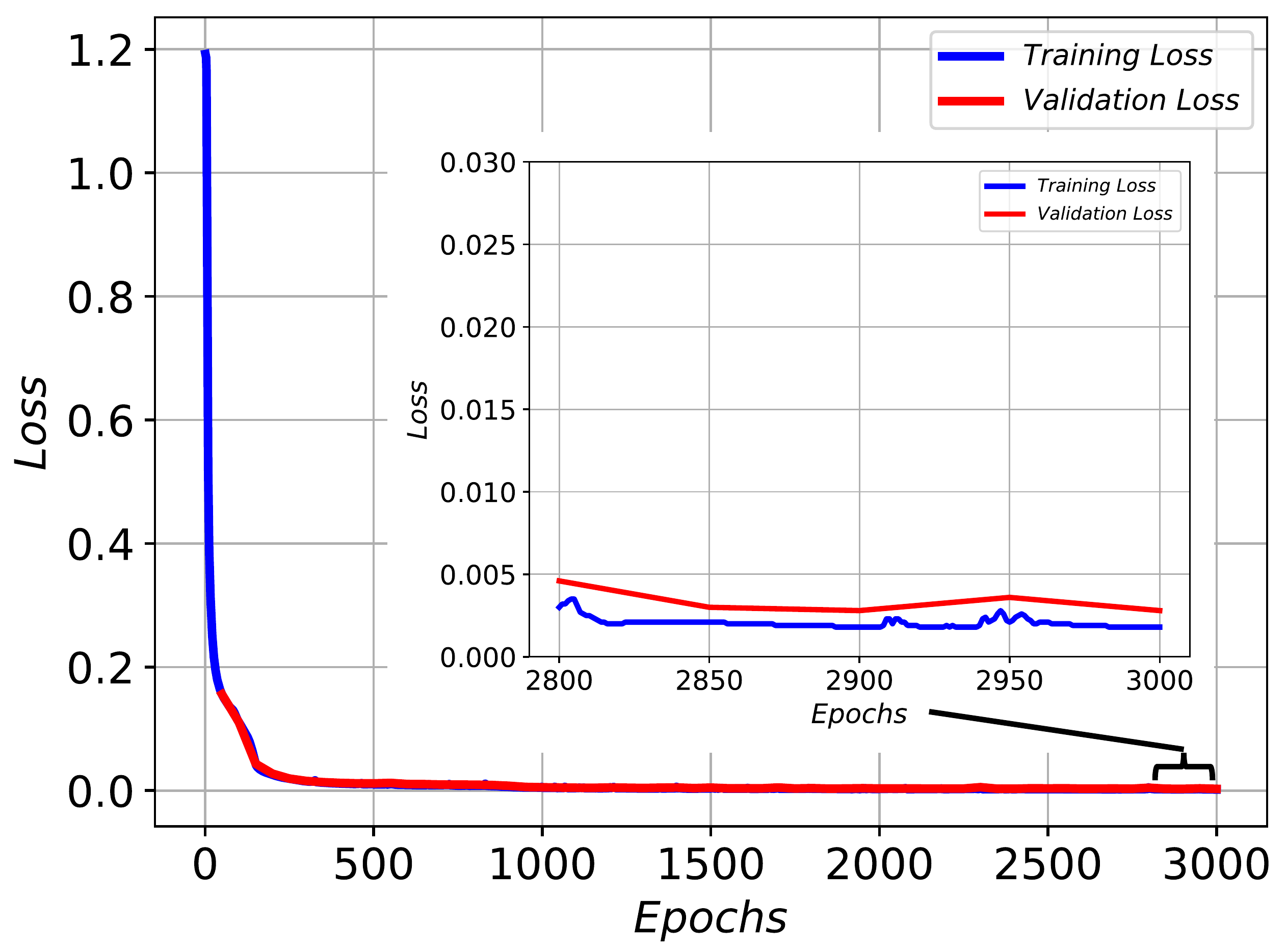}
\caption{The training loss and testing loss history for the LDM.}
\label{pitch_distb_LSTM_loss}
\end{figure}

Thus, we arrange the input training data as in condition 2 in \revision{Eq.}~\eqref{eqn:LDM_input}, and the output data as in \revision{Eq.}~\eqref{eqn:LDM_output}. We use a standard mean-squared error loss function for this training. Since each input sequence represents an individual simulation case, we have a total of 1013 sequences. In order to avoid overfitting, we randomly split this set into 700 sequences for the training dataset, and 313 sequences for the validation dataset. The training loss and validation loss are shown in \fig~\ref{pitch_distb_LSTM_loss}. For the purpose of saving computational time, the validation loss is calculated every 50 epochs. Both the training and validation loss decrease during the training, showing that the trained model generalizes well. Based on the loss histories, we use the model generated at the $3000$th epoch for the purposes of the MBRL. For reference, the LDM training in this case requires approximately $30$ minutes on an NVIDIA GTX $3070$ Ti GPU card.

\begin{figure}[hbtp!]
\centering
\begin{subfigure}[b]{0.63\textwidth}
    \includegraphics[width=\textwidth]{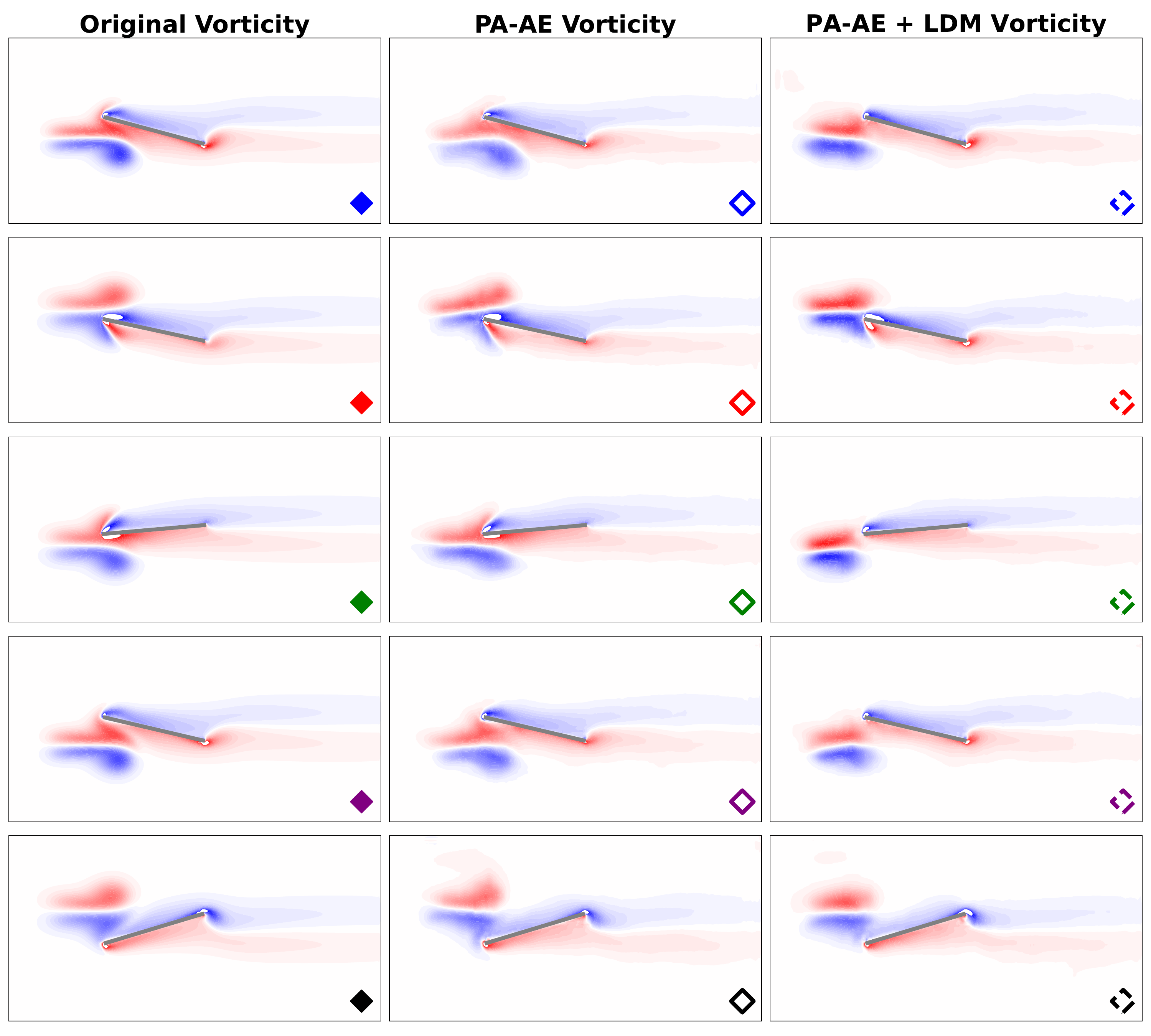}
    \caption{The predicted vorticity field comparison at $\mathbf{t^*=0.75}$}
    \label{pitch_distb_LSTM_vorticity_t25}
\end{subfigure}
\hfill 
\begin{subfigure}[b]{0.63\textwidth}
    \includegraphics[width=\textwidth]{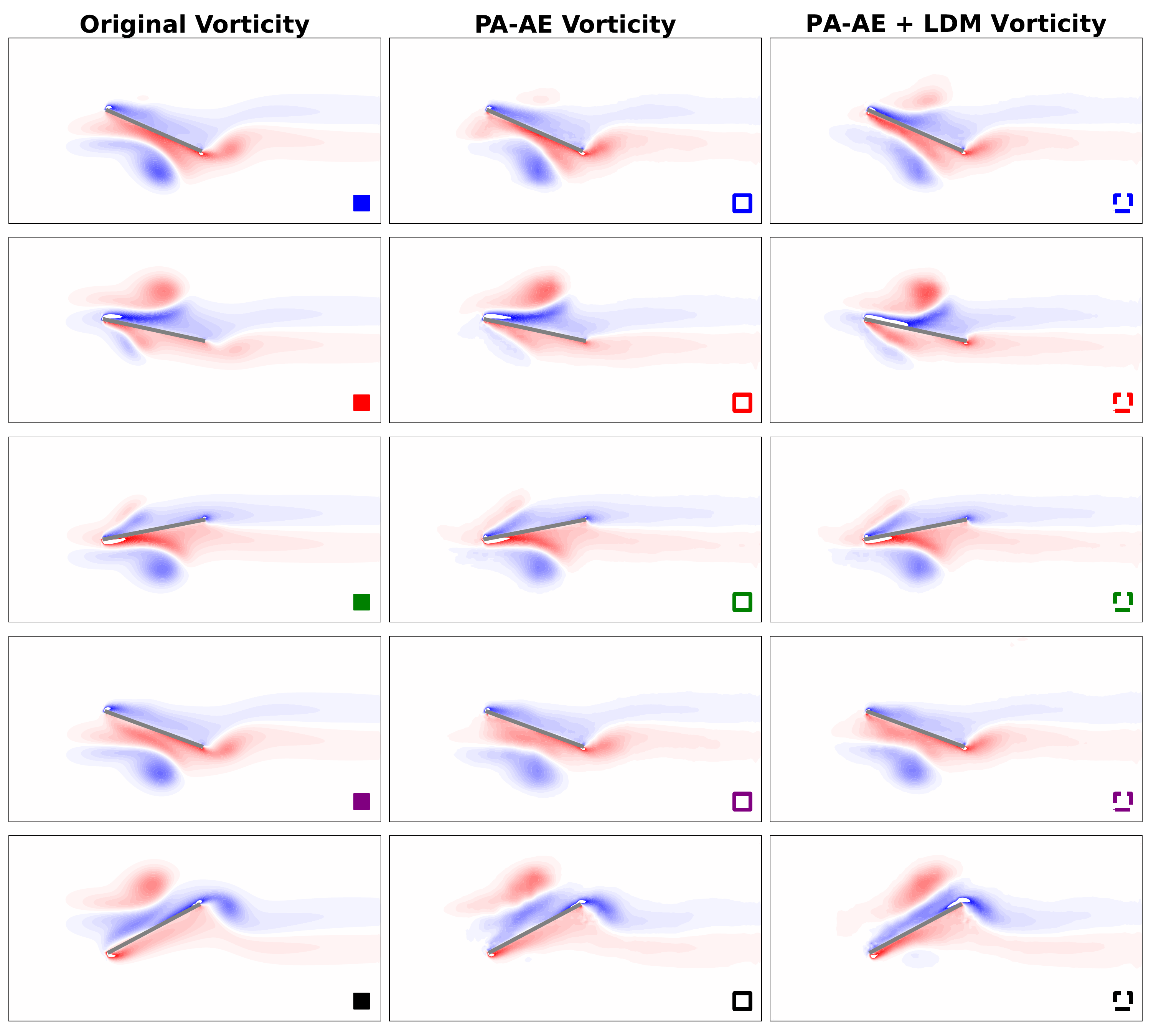}
    \caption{The predicted vorticity field comparison at $\mathbf{t^*=1.05}$}
    \label{pitch_distb_LSTM_vorticity_t35}
\end{subfigure}
\caption{Comparison among true, PA-AE, and PA-AE + LDM predicted vorticity field for 5 unseen cases labeled by different colors. The corresponding time instances corresponding to the visualization are $\mathbf{t^*=0.75}$ and $\mathbf{t^*=1.05}$. Each symbol color corresponds to a different case. Solid, blank, and dashed symbols represent the true, PA-AE, PA-AE+LDM vorticity, respectively.} 
\label{pitch_distb_LSTM_vorticity}
\end{figure}

\begin{figure}[hbt!]
\centering
\begin{subfigure}[b]{\textwidth}
    \includegraphics[width=\textwidth]{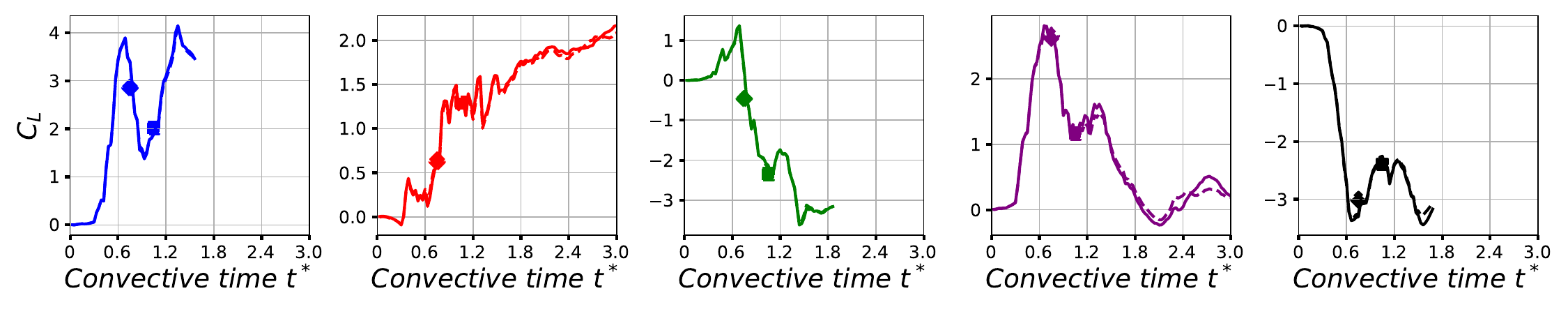}
    \caption{The lift coefficient prediction.}
    \label{pitch_distb_LSTM_lift}
\end{subfigure}
\hfill
\begin{subfigure}[b]{0.6\textwidth}
    \includegraphics[width=\textwidth]{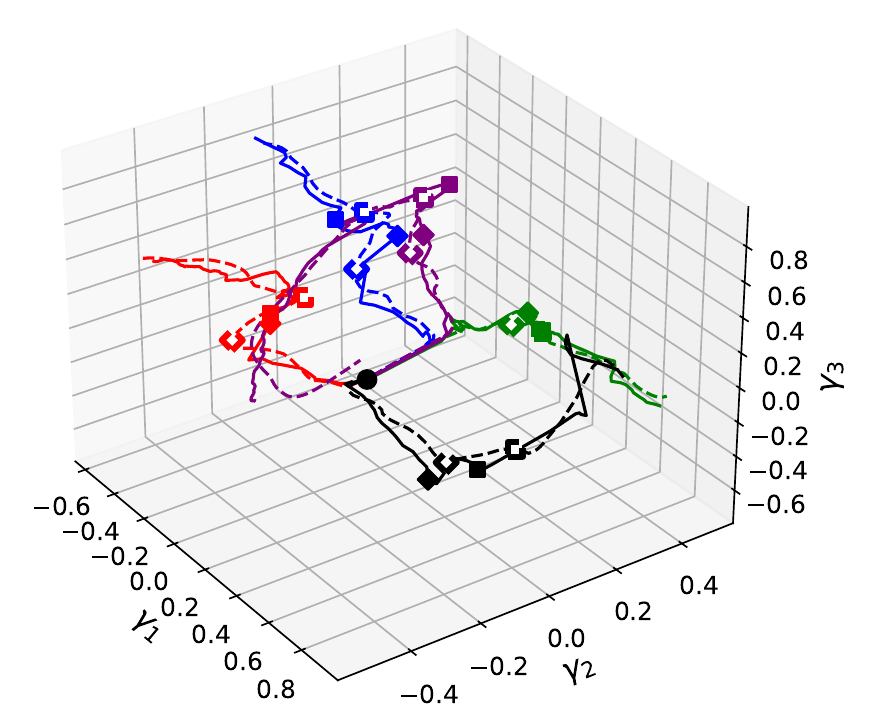}
    \caption{The latent variables $\boldsymbol{\gamma}_k=(\gamma_{1_k},\gamma_{2_k},\gamma_{3_k})$ prediction. All trajectories start from the symbol $\blackcirc$.}
    \label{pitch_distb_LSTM_latent}
\end{subfigure}
\caption{The lift prediction and the corresponding latent variables prediction for the same 5 unseen cases as \fig~\ref{pitch_distb_LSTM_vorticity} with the same convention of color and symbol as in \fig~\ref{pitch_distb_LSTM_vorticity}. The solid line \protect\tikz[baseline]{\protect\draw[solid, thick] (0,.8ex) -- (0.3,.8ex);} represents the ground truth, and the dashed line \protect\tikz[baseline]{\protect\draw[dashed, thick] (0,.5ex) -- (0.3,.5ex);} represents the prediction from LDM.}
\label{pitch_distb_LSTM_lift_latent}
\end{figure}

Before we employ the LDM model for the MBRL, it is necessary to see how the LDM performs in predicting the vorticity field and lift coefficient, especially for the period of strong gust-plate interaction. Accurate prediction is crucial since we will depend strongly on the accuracy of predicted latent variables to train the MBRL in this three-dimensional latent space. In \fig~\ref{pitch_distb_LSTM_vorticity}, we perform a comparison for five unseen cases between the true vorticity field, the PA-AE reconstructed vorticity field decoded from the true latent variable $\boldsymbol{\gamma}_k$, and the LDM + PA-AE predicted vorticity field, decoded from the LDM-predicted latent variable $\boldsymbol{\hat{\gamma}}_k$. It is apparent that the LDM actually can predict the fluid dynamics well, including the configuration of the boundary layer to identify the angle of attack of the flat plate, the variation of the gust structure, and the wake behavior, albeit with some visible errors in predicting the position of the gust, particularly at time $t^*=0.75$. Additionally, the lift coefficient prediction is depicted in \fig~\ref{pitch_distb_LSTM_lift}. We can see that the lift prediction is very accurate even for time intervals when the strong gust-plate interactions happen. It is interesting to note that the LDM can predict the lift coefficient well even when there is rapid pitch oscillation of the flat plate, as indicated in the case with the red curve. This agreement underscores the robustness and the good generalization of the LDM model. We also observe that the lift prediction may have some bias at large $t^*$, such as the time period $t^* \in [2.1, 3.0]$ in the case denoted by a purple curve. This error arises from the aforementioned dearth of cases in the training dataset that proceed to 3 full convective times. Consequently, the LDM demonstrates slightly weaker performance in predicting dynamics at larger values of $t^*$ compared to its predictions at smaller $t^*$. This could be improved by increasing the number of training cases, or by changing the strategy for generating the training data, such as reducing the magnitude of the $\ddot{\alpha}$ in the random policy. However, this latter choice must be balanced with the need to cover a sufficiently large action space to obtain a good RL policy.

\fig~\ref{pitch_distb_LSTM_latent} also compares the prediction of the latent variable trajectories with the truth trajectories. The latent variables are predicted well for all five unseen cases. However, we can observe the same slight degradation at later times as we observed with the lift prediction; this is most apparent in a relatively large bias in the last portion of the purple trajectory. However, we can also observe that the sensitivity of the mapping between the latent space and lift is not uniform. For instance, although there is bias error in the last latent point of the purple trajectory, the lift coefficient prediction for the corresponding point is very close to the true value.  On the other hand, although the prediction of the very last point of the red trajectory is very close to the true latent variable, the lift coefficient prediction of the corresponding point shows a relatively large error. This underscores the need to further explore the mapping to and from the latent space. However, based on the generally strong performance of the prediction, the LDM model is accurate and sufficiently robust to predict a wide variety of complex and large-amplitude flows in the class of disturbances and action sequences covered by the interpolation range of the training dataset. Therefore, we can employ it with confidence for the MBRL study in Section~\ref{sec:MBRL_realization}.

\section{MBRL Realization}\label{sec:MBRL_realization}
In this section, we employ the trained PA-AE and LDM to perform MBRL in the gust-airfoil flow environment considered in Section~\ref{sec:model_application}. The objective of this MBRL is to find a policy for pitch accelerations that minimizes the variation of the lift coefficient $C_{L_k}$ about a reference lift coefficient $C_{L,\text{ref}}=0$. In other words, in this study, we aim to pitch the airfoil to make $|C_{L_k}|$ remain as close to $0$ as possible during a gust interaction. Consequently, we define the reward function as follows: 
\begin{equation}\label{eqn:reward}
    R_{k+1} = \frac{1}{|C_{L_{k+1}}|/C_{L,\text{scale}}} - \left(\frac{|C_{L_{k+1}}|}{C_{L,\text{scale}}}\right)^2 - 2 \frac{|\ddot{\alpha}_{k}-\ddot{\alpha}_{k-1}|}{\ddot{\alpha}_{\text{scale}}} - \begin{cases} 
    100 & \text{if truncated} \\
    0  & \text{if not truncated}
  \end{cases}
\end{equation}
where the $C_{L,\text{scale}}=0.1$ and $\ddot{\alpha}_{\text{scale}}=10$. This reward definition is inspired from the earlier work of Beckers and Eldredge \cite{beckers2024deepreinforcementlearningairfoil}, with some notable exceptions. The leading term in their reward consisted of a constant value, rewarding the agent for each action step in which the episode continues without early termination. However, we have observed that the RL agent struggles to effectively learn a policy to manage scenarios where the variation of the lift coefficient magnitude $|C_{L_k}|$ is small without any active control, such as when the gust has moved past the flat plate. Thus, we amplify this positive reward with the inverse term. With the second term, we find that the RL agent learns a policy that can effectively reduce the high variation of the $C_{L_k}$ for the scenarios in which the gust is strongly interacting with the airfoil. The purpose of the third term is to prevent the flat plate from undergoing impractically rapid pitch oscillations. Finally, since we truncate the model training simulations if $|\alpha_k| \geq 45^\circ$, we use the same constraint for the RL training. If an episode is truncated, then a relatively large penalty is given so that the RL agent learns this constraint effectively. To avoid some of the issues with model accuracy encountered at later times, as described in Section~\ref{sec:pitch_distb_LDM}, we only train the RL agent over the first $2.1$ convective time units, i.e., the first $70$ action steps. 

\begin{figure}[hbt!]
\centering
\includegraphics[width=0.6\textwidth]{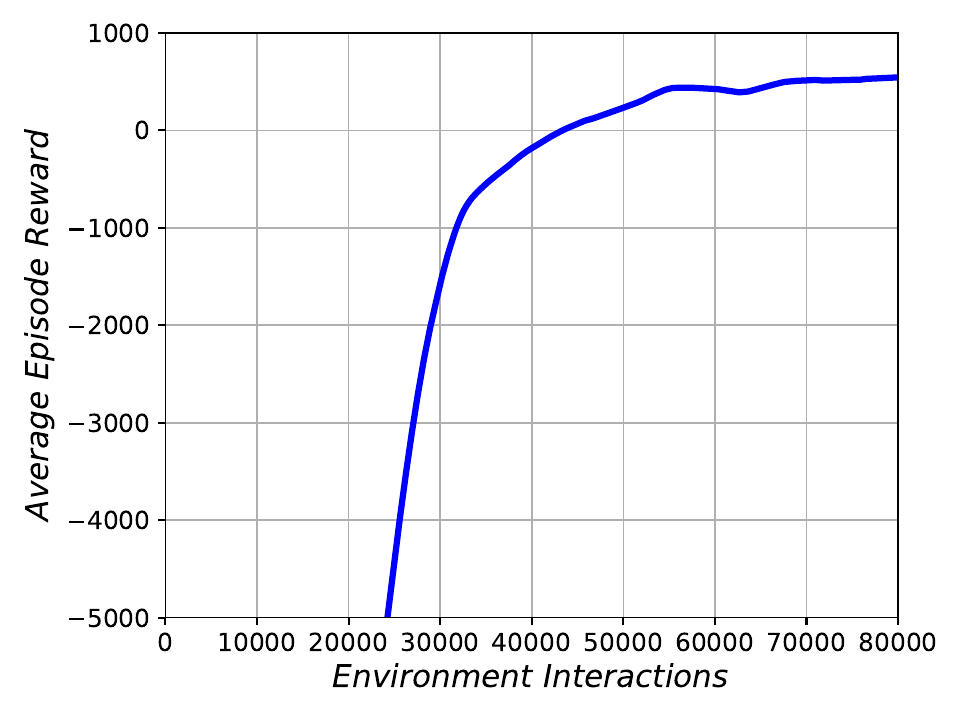}
\caption{Average episode reward during RL training. Rewards are averaged every four episodes and recorded at each checkpoint. A moving average with a window size of 50 checkpoints further smooths the learning curve. Only data points where the average episode reward exceeds -5000 are displayed to maintain an appropriate plot scale.}
\label{pitch_distb_RL_reward}
\end{figure}

The training process is entirely carried out in the three-dimensional latent space, as illustrated in \fig~\ref{RL_Overview}. The observation here is defined as $O_{k+1}=(\boldsymbol{\hat{\gamma}}_{k+1},C_{L_{k+1}},\alpha_{k+1},\dot{\alpha}_{k+1})$. The maximum action magnitude is $\ddot{\alpha}_{\text{max}}=10$, which also bounds the action space, so that the RL agent is constrained to learn a policy within this constraint. We trained the RL agent over $80000$ interactions by using the TD3 algorithm, implemented in the Stable Baselines3 library \cite{JMLR:v22:20-1364}. The initial state is the same for each training episode, and the auxiliary parameters $\mathbf{P}$ describing the gust are randomly chosen. The learning curve, shown in \fig~\ref{pitch_distb_RL_reward}, clearly shows that the agent learns to mitigate the variation of the lift coefficient significantly during the training process, and the average episode reward converges after around $70000$ interactions. For reference, the RL training up to $80000$ interactions within the latent space environment requires $\mathcal{O}(10)$ minutes, while the conventional RL training within the CFD environment requires $\mathcal{O}(1)$ day, on an Apple M1 Pro chip with a single CPU core.

Once training is deemed completed (at $80000$ interactions), we evaluate the policy both in the latent space environment and, crucially, in the true (CFD) environment. In the latent space environment for a given gust scenario, we obtain a sequence of actions from the trained policy; in the true environment, we apply this same action sequence for the same gust scenario, for comparison with the latent space results. It is also important to compare the trained policy's performance with that of no control. Therefore, in the results that follow, we make this comparison in each of the two environments. For these purposes, we use four representative cases with the gust parameters detailed in Table \ref{table:gust_parameters}.

\begin{table}[bt!]
\caption{\label{table:gust_parameters} The Gaussian parameters of the gust in evaluation cases}
\centering
\begin{tabular}{lcccccc}
\hline
Case ID& $D_y$ & $\sigma/c$ & $y_0/c$ \\\hline
A& -0.78405& 0.08305& 0.00859 \\
B& 0.72003& 0.09386& -0.12283 \\
C& 0.33421& 0.08221& 0.15137 \\
D& -0.53989& 0.08411& 0.19080 \\
\hline
\end{tabular}
\end{table}

\begin{figure}[hbt!]
\centering
\includegraphics[width=1.0\textwidth]{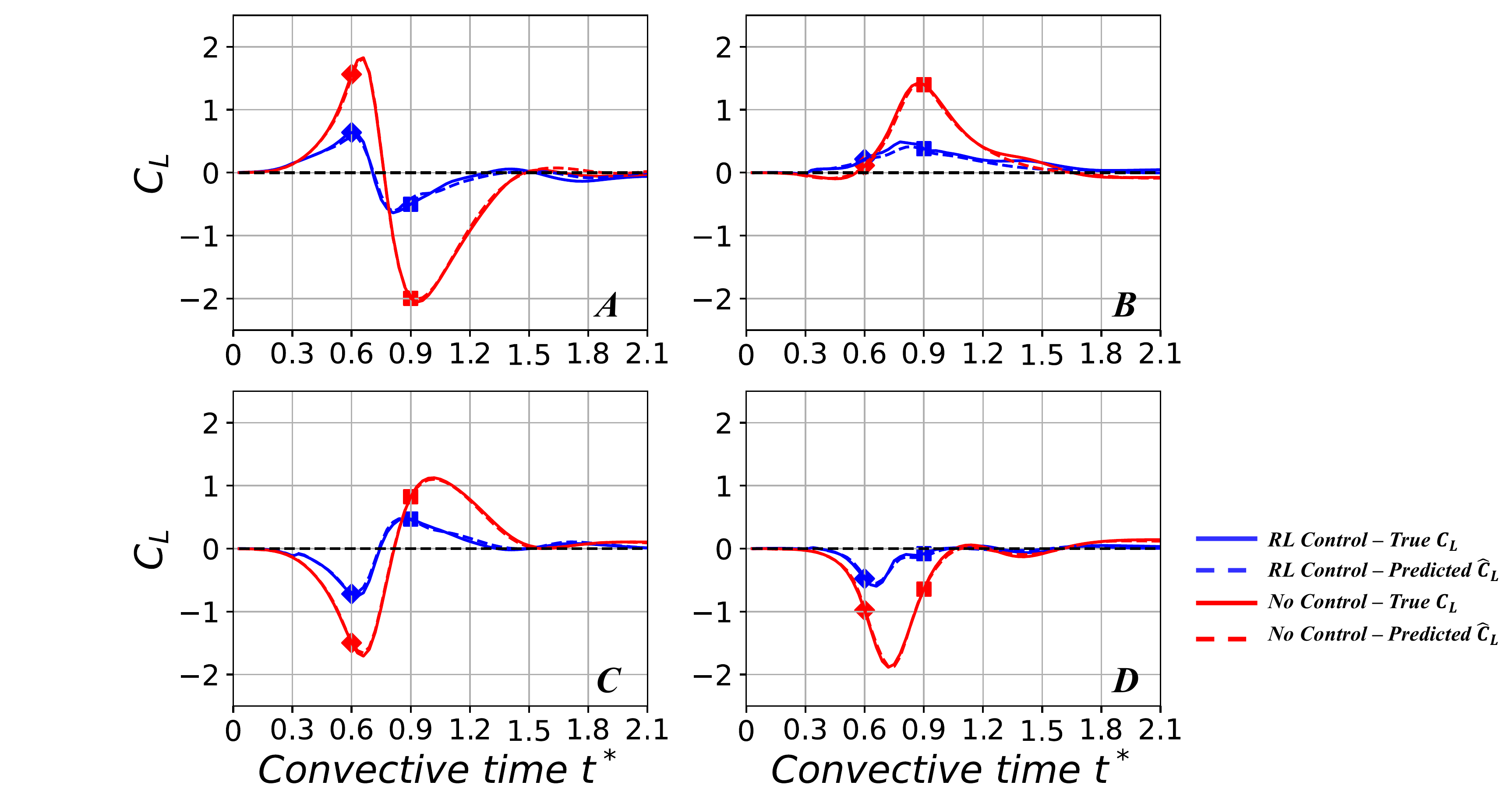}
\caption{Examples for the lift comparison between the RL control and no control. $\mathbf{A},\mathbf{B},\mathbf{C},\mathbf{D}$ represent four different evaluation environments with gust for which the parameters are randomly-generated. The true $C_L$ is evaluated in the true environment (CFD simulation). The predicted $\hat{C}_L$ is evaluated in the LDM environment.}
\label{pitch_distb_RL_lift}
\end{figure}

The four representative cases elicit four distinctly different types of variation of lift coefficient for the flat plate airfoil without control, as depicted in \fig~\ref{pitch_distb_RL_lift}. Indeed, we have found that these cases represent the only types of uncontrolled lift variation for this class of gusts, and other responses are only different by a change of magnitude or phase. This observation underscores the trained policy's strong generalization capabilities, as it consistently outperforms the no-control policy across all scenarios. Specifically, the trained policy is adept at managing not only scenarios with intense gust-plate interactions but also situations where the lift coefficient remains relatively steady even in the absence of active control. Furthermore, it is clear that the results from evaluation in the true environment align very closely with those in the latent space environment, instilling further confidence that the trained LDM model serves as a suitable surrogate for RL training. The success of the RL agent in mitigating the lift variation is due in part to the physical fidelity of the model on which it is trained. This accuracy ensures that the observation and reward received at each time by the agent include the correct information about the flow and about the action's effect on it, so the agent is able to learn an appropriate policy effectively.

\begin{figure}[hbt!]
\centering
\includegraphics[width=1.0\textwidth]{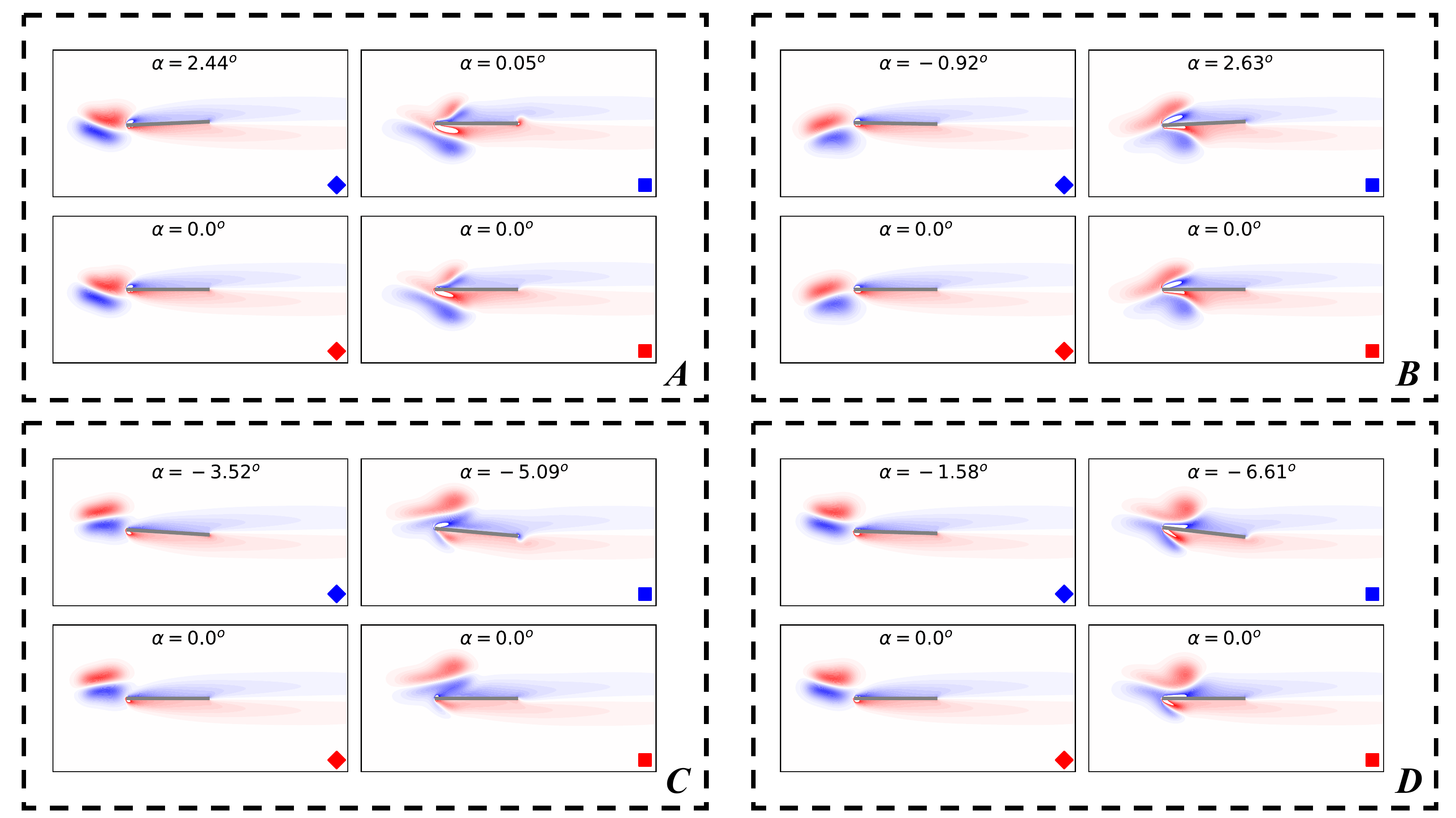}
\caption{Visualization of the true vorticity field labeled by $\Diamond$, and $\square$ corresponding to the same time instance indicated by the same symbol in \fig~\ref{pitch_distb_RL_lift}, and following the same convention for use of symbol color: blue represents the RL control case, and red represent the no control case. The corresponding pitch angle $\alpha$, measured in degrees and positive in the counter-clockwise direction, is included in each of the subfigures.}
\label{pitch_distb_RL_vorticity}
\end{figure}

\begin{figure}[hbt!]
\centering
\includegraphics[width=1.0\textwidth]{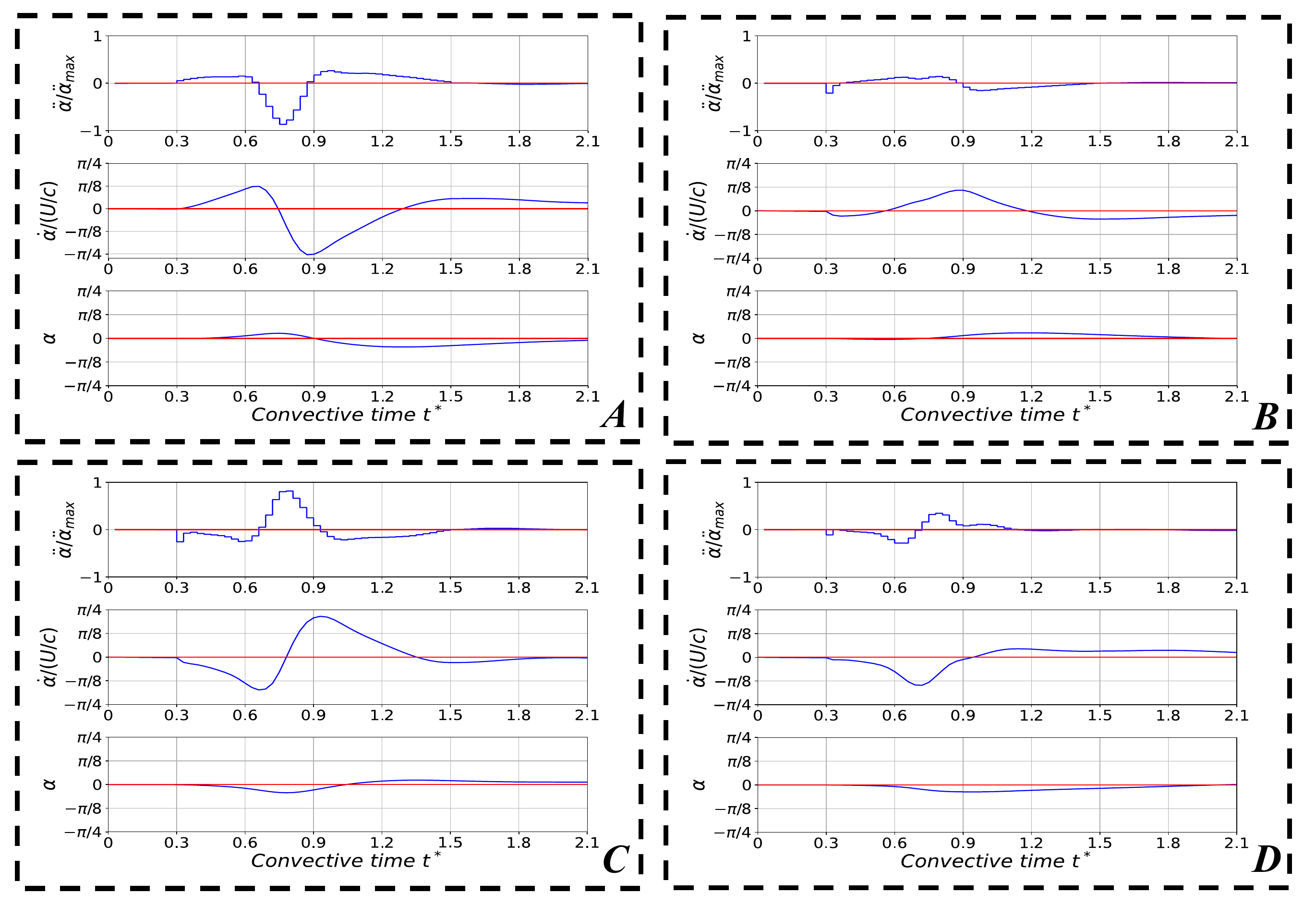}
\caption{Visualization of the action $\ddot{\alpha}$, angular velocity $\dot{\alpha}$, and the pitch angle $\alpha$ for the four representative cases. With the same convention of color use as in \fig~\ref{pitch_distb_RL_lift}: blue represents the result of RL control, and red represents the results of no control.}
\label{pitch_distb_RL_action}
\end{figure}

The true vorticity fields corresponding to $t^*=0.6$ and $t^*=0.9$ for each of the four cases, with and without control, are visualized in \fig~\ref{pitch_distb_RL_vorticity}. We can see that the gust interacts strongly with the leading edge of the flat plate, and these interactions are further influenced by the pitching motion of the plate, so the flow exhibits a strong non-linearity in this combination of gust and action. However, even in this complex environment, the RL agent effectively learns the underlying mechanisms for mitigating lift variation through control of the pitching motion. Furthermore, the corresponding profiles of the $\ddot{\alpha}$, $\dot{\alpha}$, and $\alpha$ for the four representative cases are shown in \fig~\ref{pitch_distb_RL_action}. We can observe that the agent not only learns how to mitigate the lift variation, but also learns to avoid a rapid oscillatory control input because of the second term in the reward function defined in \revision{Eq.}~\eqref{eqn:reward}.

\section{Conclusions}
In this study, we have proposed a model-based reinforcement learning (MBRL) strategy for unsteady aerodynamics applications by introducing a novel reduced-order modeling framework for the disturbed flow. As far as the authors know, this is the first effort to apply model-based reinforcement learning for flow control of unsteady aerodynamics. The framework consists of a physics-augmented autoencoder (PA-AE) and a latent dynamics model (LDM). We first demonstrated that the PA-AE, inspired from the work of Fukami and Taira \cite{Fukami2023}, can effectively reduce the $\mathcal{O}(10^4)$ dimension of the CFD space to a three-dimensional latent space while preserving the main flow structures and key physical variable information. Then, we showed that the LDM is capable of predicting the latent dynamics and the physical variables for long times, given only the initial condition, the corresponding action sequences, and some parametric information about the disturbance. We confirmed the \revision{accuracy and }robustness of the modeling framework by applying it \revision{to the scenario of} a pitching airfoil in a highly disturbed environment. Furthermore, after wrapping the model as the RL environment, we realized the MBRL by controlling the pitching of an airfoil to mitigate the lift variation in a highly-disturbed flow environment.

We introduced the disturbance (gust) by applying localized Gaussian forcing to the flow environment. The specific gust forcing parameters---the amplitude, the size, and the vertical location---were randomly sampled from case to case. The resulting flow response to the gust and the pitching action exhibited strong nonlinear dependence on the parameters. We showed that the PA-AE could capture the main information of the flow, including the gust structure, boundary layer, and the wake, by augmenting the autoencoder with the lift coefficient during the training. We found that, by incorporating some auxiliary parameters about the gust into the input, the LDM could effectively distinguish between cases and predict the latent dynamics for up to $100$ action steps. \revision{Additionally, a study on a VAWT problem, as presented in Appendix, confirms the generalizability of our approach by showing that autoencoder augmentation can also be applied to the power coefficients and that the auxiliary parameters are not necessary in a disturbance-free environment.}

Finally, we incorporated the PA-AE and LDM into the RL environment and realized the MBRL. Specifically, we have tasked the MBRL with learning a policy that would pitch an airfoil to mitigate the lift variation during random gust encounters. We showed that the trained RL agent successfully met this objective, achieving significantly smaller lift variation compared to the uncontrolled case in a wide variety of gust conditions. Furthermore, by evaluating the policy in both the model environment and the true flow environment, we found that the policy resulted in nearly identical lift coefficient histories and vorticity fields, underscoring the accuracy of the underlying model for the duration of the RL episode.

The current study proves the feasibility of using MBRL in a highly-disturbed flow environment with strong nonlinear flow physics. However, there are many ways in which this study can be expanded in future work. First, though the overall training time---accounting for PA-AE, LDM, and the RL itself---required for this study does not exhibit a clear advantage over the model-free RL approach followed by Beckers and Eldredge \cite{beckers2024deepreinforcementlearningairfoil}, the model-based approach has an important advantage: the time required to train a new MBRL policy (for example, for different reward definition or a different sampling of disturbances) is negligible (a few minutes), whereas every policy training of the model-free approach requires approximately the same large effort (around a day for this 2D problem). Furthermore, we have not been particularly economical with the most expensive part of the training in the current study, since we have generated a set of CFD training data by following a random policy. This approach appears to adequately cover the space of possible states and actions for the eventual RL training. However, it would be useful to investigate this further, so that computational (or experimental) effort is not wasted on superfluous cases, possibly by following a more strategic approach. For example, one might use a MBRL paradigm in which the model is trained simultaneously with the policy, so that favorable states and actions are weighted more heavily in interactions with the full environment for the model and policy training. Also, we have not thoroughly investigated the robustness or optimality of the RL policy in this work, but merely sought to demonstrate that a model-based approach is feasible. Thus, more research should be pursued to study the RL policy, similar to the investigation of model-free RL of Beckers and Eldredge \cite{beckers2024deepreinforcementlearningairfoil}. 

Finally, we have not considered practical sensing in the RL component of this study. Rather, in our policy training, we have assumed that we had access to the full flow field (or, more specifically, the latent state), along with the lift and the airfoil configuration, when choosing the action. However, in a practical scenario, this knowledge of the flow state would only be available indirectly (and perhaps only partially) via sensor measurements, e.g., surface pressure sensors, and it remains an open question whether the state is observable from such surface pressure data (i.e., whether the sensor data uniquely map to a latent state trajectory). Beckers and Eldredge \cite{beckers2024deepreinforcementlearningairfoil} showed that an effective policy could be trained to rely on a set of surface pressure sensors, provided that a short history of the sensor data was retained to mitigate the effects of partial observability. In the model-based framework, we would need to learn the mapping from the sensor measurements to the latent state. Furthermore, for practical use in a real (physical) scenario, the sensitivity of the mapping to sensor noise should be accounted for, e.g., by wrapping the model in an ensemble filtering framework. 

\setcounter{figure}{0}  
\renewcommand{\thefigure}{A\arabic{figure}}  
\section*{\revision{Appendix A: Vertical-Axis Wind Turbine in a Disturbance-Free Environment}}
In this appendix, we investigate another application of our proposed PA-AE and LDM framework: a two-dimensional representation of a vertical-axis wind turbine (VAWT) in a disturbance-free environment. The purpose of this example is to explore whether the modeling framework is capable of adapting to broader scenarios, including those involving more complex airfoil motions.
\section*{A.1. Problem Set-Up and Simulation}
\begin{figure}[hbt!]
\centering
\includegraphics[width=0.6\textwidth]{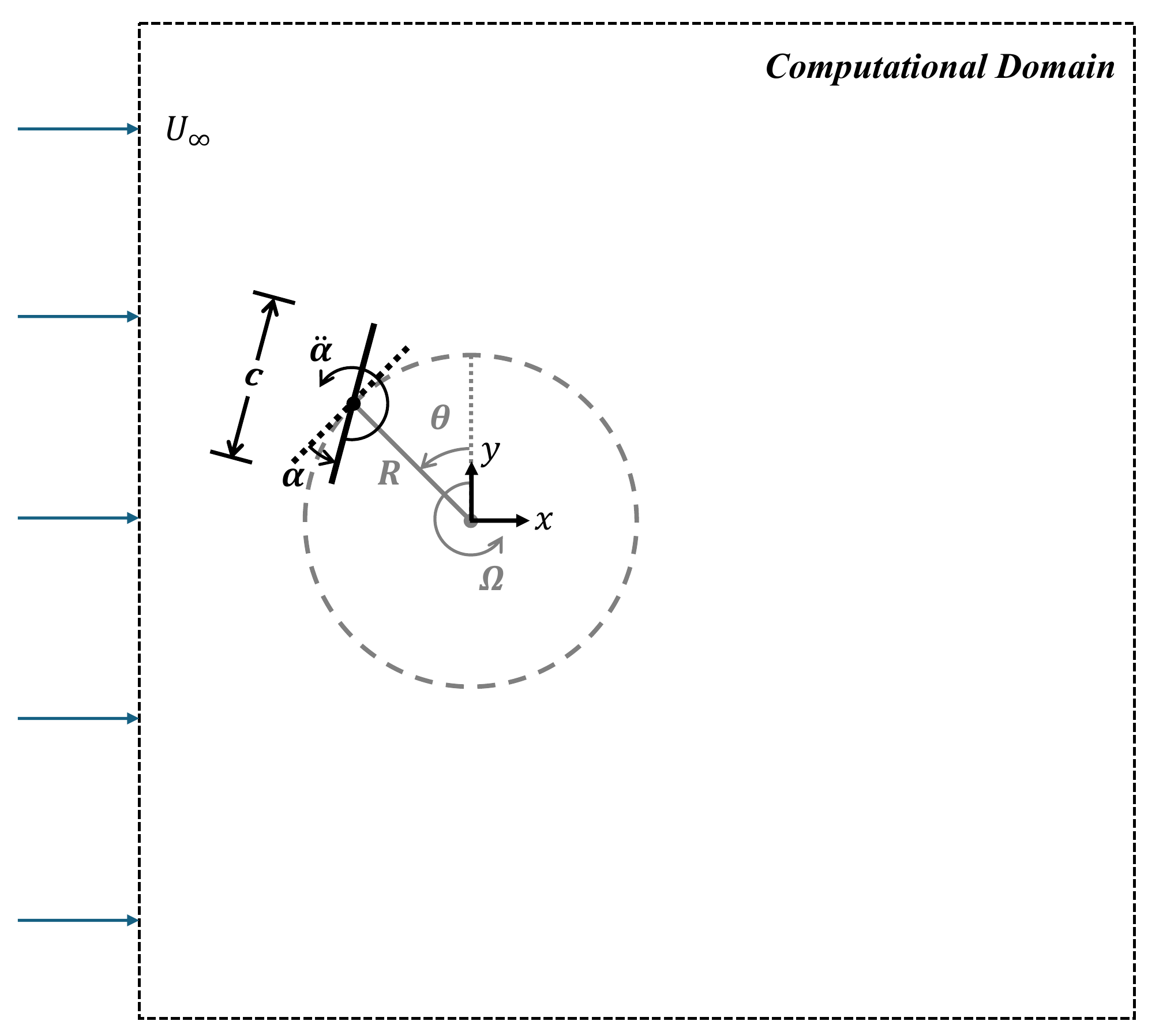}
\caption{The schematic figure of the problem set-up for the VAWT in the gust-free environment}
\label{VAWT_schematic}
\end{figure}

In this problem, we consider a flat plate airfoil of chord length $c$ performing two prescribed motions simultaneously, as depicted in \fig~ \ref{VAWT_schematic}. The first motion consists of the pivot point at the midchord of the flat plate rotating around a central axis at a distance $R$ with a constant counter-clockwise angular velocity $\Omega$. The second motion involves the pitching about the pivot point with prescribed, random time-varying angular acceleration $\ddot{\alpha}$, with $\alpha$ initially zero and measured counter-clockwise from the axis perpendicular to the radius connecting the plate to the central axis. There is a uniform flow with constant horizontal velocity $U_{\infty}$. The pitching motion will be referred to as the action in this problem, as it has been identified as a means of control in the presence of wind speed variations \cite{le2024optimal}. For the case considered in this work, we set the rotating radius $R=c$ and the rotating angular velocity $\Omega=0.8U_{\infty}/c$, so we have a constant tip-speed ratio $\Omega R/U_{\infty}=0.8$. The Reynolds number is set at $Re = U_\infty c/\nu = 100$. As in the previous example, time is non-dimensionalized by $c/U_\infty$ and spatial coordinates by $c$. Thus, the time to complete one revolution about the axis is $T_p^* = 5\pi/2$. For each simulation we perform in this study, the action sequence will be randomly chosen, and for model training purposes we will save the resulting vorticity field snapshot $\omega_k$, the action $A_k$, and the power coefficient $C_{p_k} = \dot{W}_k/(\frac{1}{2}\rho U_{\infty}^3 c)$ with $\dot{W}_k$ representing the net power extracted from the flow by the plate, at each action step $k$.

In the simulation, we use the same flow solver as described in Section~\ref{sec:pitch_distb_set_up} to solve the flow. We set the computational domain over $[-1.98,3.98]\times[-2.98,2.98]$ with the rotational axis at the origin and employ a uniform Cartesian grid with grid size $\Delta x^* = 0.02$, so that we have a grid of $300 \times 300$ points in the computational domain. We set the time step size as $\Delta t^* = 5\pi/2000$, corresponding to 1000 time steps per revolution. Before we collect any data, we conduct one simulation in which the flat plate completes one revolution without any pitching motion to ensure there is already some wake when we start to collect the data. We then carry out a second revolution in which we restrict the prescribed $\ddot{\alpha}$ to be piecewise constant over intervals of $5\pi/200$ convective time units so that the maximum number of action steps for each simulation case is $100$. We save the data at each action step, so that there are at most 101 vorticity snapshots, 101 power coefficient data points, and 100 action data points saved for each simulation case. Similar to the previous example, we terminate the simulation case early if $|\alpha|\geq 90^\circ$. The actions are drawn from a policy $\ddot{\alpha}=0.5\text{randn()}$, which balances the need for training data over an entire rotation and a wide range of angular motion amplitudes. Ultimately, we perform 415 simulation cases, of which around 400 simulation cases are not truncated. It should be stressed that the customized random policy we have implemented in this work does not enforce periodic motion, which means that the initial pitching angle, $\alpha_{k=0}$, and the final pitching angle, $\alpha_{k=100}$ are generally not equal. If we want to enforce periodicity, we could instead represent the pitching angle $\alpha$ as a Fourier series with harmonics of the angular velocity $\Omega$. This approach introduces an intriguing optimization problem regarding the power coefficient under conditions of periodicity. However, it falls outside the scope of the current study, for which our primary objective is to demonstrate the robust generalizability of our proposed PA-AE and LDM framework in scenarios involving complex airfoil motions.

\begin{figure}[hbt!]
\centering
\includegraphics[width=1.0\textwidth]{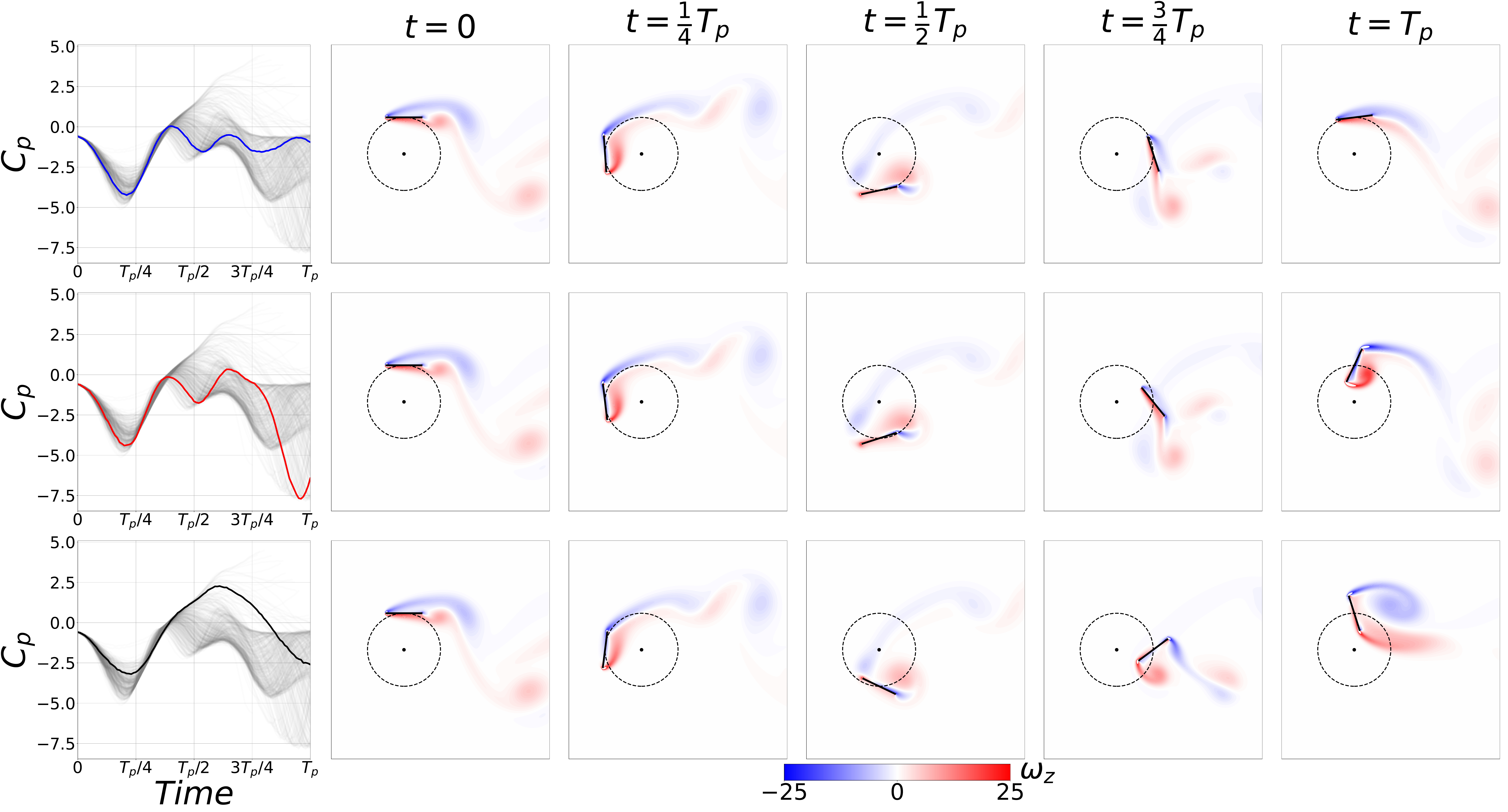}
\caption{Three representative examples of the power coefficient response and the corresponding vorticity fields are shown. The power coefficient curves for each of the three cases are highlighted by different colors. The vorticity field is visualized at each quarter of a period for each case. The grey color power coefficient curves represent all the power coefficient data obtained in this study.}
\label{VAWT_original_data}
\end{figure}

The power coefficients and visualization of the vorticity fields for three representative cases are depicted in \fig~ \ref{VAWT_original_data}. Since there is no external disturbance in this problem, the main differences in both the power coefficient and the vorticity field arise from the different variation of the pitching angle $\alpha$. The first example, denoted by blue, has a relatively small $\alpha$ variation, but the second and the third example, colored by red and black respectively, show a high $\alpha$ variation in opposite directions. Given the substantial variations in the pitching angle $\alpha$, the wake patterns exhibited by each case may differ significantly. Furthermore, the pronounced flow separation presents an additional challenge for our proposed model, requiring it to effectively learn not only the wake dynamics but also to accurately predict instances of flow separation.

\section*{A.2. PA-AE Training and Validation}
\begin{figure}[hbt!]
\centering
\includegraphics[width=0.5\textwidth]{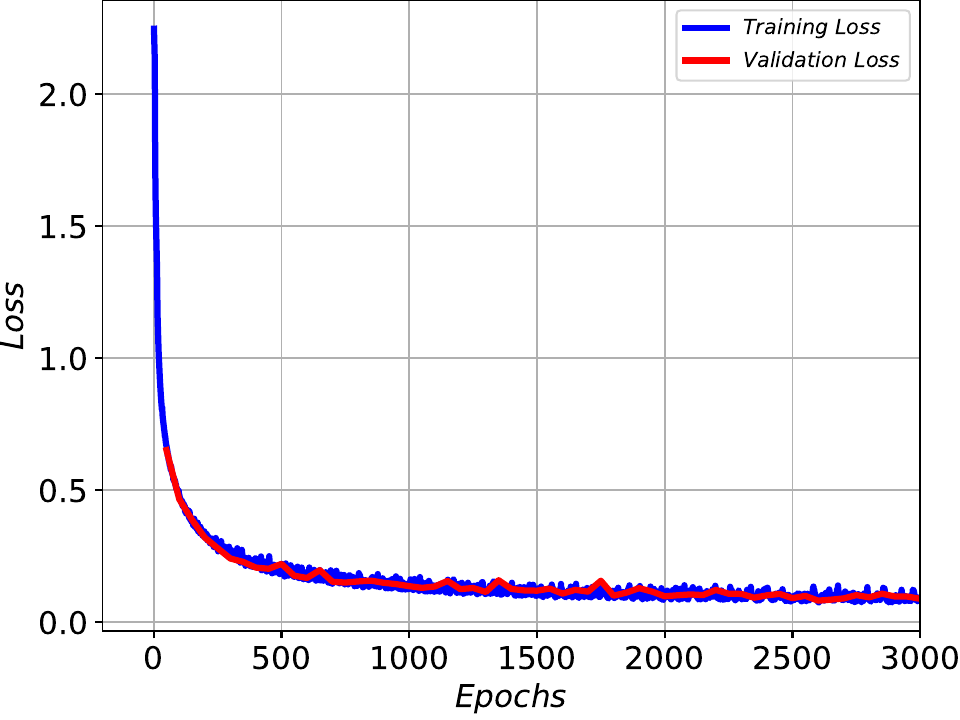}
\caption{The training loss and validation loss history for PA-AE in VAWT problem.}
\label{VAWT_AE_loss}
\end{figure}

After performing the simulations, we finally obtain a dataset consisting of $41500$ vorticity snapshots and the corresponding power coefficient data points. In order to avoid the overfitting, we randomly split this into a training dataset of $70\%$ and a validation dataset of $30\%$. The loss histories during the PA-AE training is presented in \fig~\ref{VAWT_AE_loss}. For reference, the PA-AE training requires approximately $40$ hours on an NVIDIA A100 GPU card in this problem. Then, \fig~\ref{VAWT_AE_vorticity} shows the comparison between the reconstructed vorticity and the true one. We observe that the PA-AE not only successfully reconstructs the wake but also captures the location and configuration of the flat plate by reconstructing the boundary layer structure and the flow separation structure. The power coefficient reconstruction results are depicted in \fig~\ref{VAWT_AE_power}. From this figure, it is evident that the previously observed issue of poor reconstruction towards the end of the simulation is not present in this example, as relatively fewer cases are truncated. However, we still can observe slightly more error in the reconstruction during the second half period compared to the first one. This discrepancy arises because $\ddot{\alpha}$, implemented with a small amplitude, results in minimal variation in the pitching angle during the first half of the period. Consequently, the smaller interpolation range during this early interval facilitates easier and more accurate reconstruction.

\begin{figure}[hbt!]
\centering
\includegraphics[width=0.89\textwidth]{VAWT_AE_vorticity.pdf}
\caption{The representative examples of the PA-AE vorticity reconstruction corresponding to the same three cases shown in \fig~\ref{VAWT_original_data} with the same convention of color use. The solid and blank $\circ$, $\Diamond$, $\square$, $\triangle$, and \mystar symbols represent the different time instances corresponding to the subtitles of each column of the figure. All solid symbols represent the true vorticity fields, and the blank symbols represent the reconstructed ones.}
\label{VAWT_AE_vorticity}
\end{figure}

\begin{figure}[hbt!]
\centering
\includegraphics[width=1.0\textwidth]{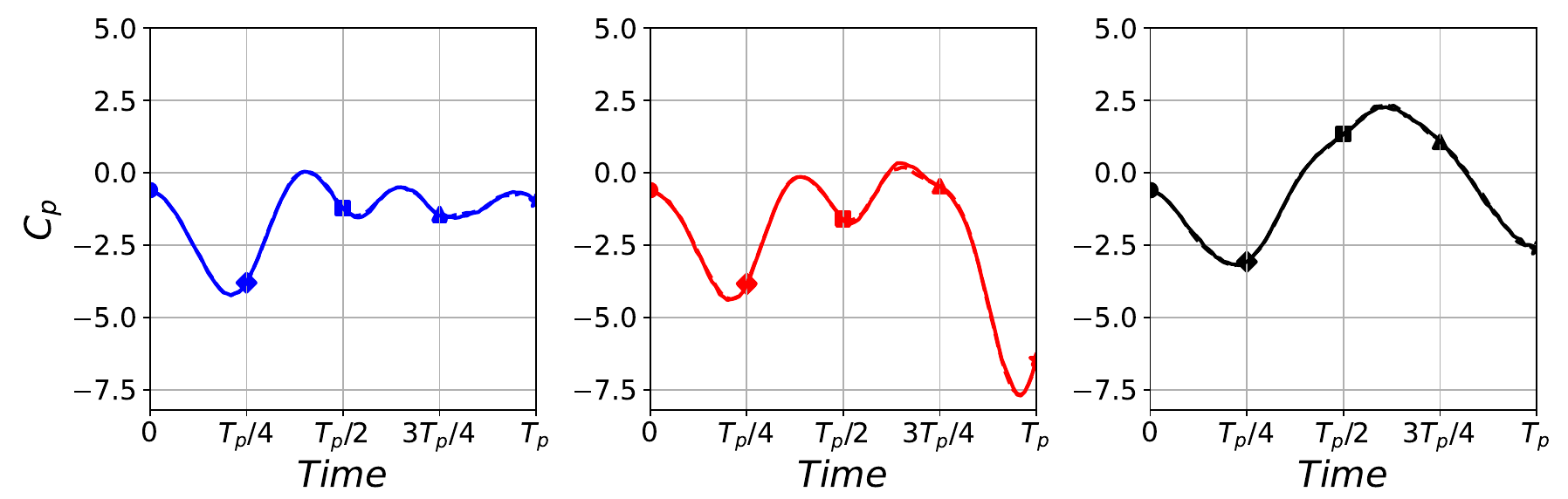}
\caption{The power coefficient comparison between the truth and the reconstructed one from PA-AE. The same three cases with the same convention of color use and symbol use as \fig~\ref{VAWT_AE_vorticity} are shown here. The solid line \protect\tikz[baseline]{\protect\draw[solid, thick] (0,.8ex) -- (0.3,.8ex);} represents the true power coefficient, and the dashed line \protect\tikz[baseline]{\protect\draw[dashed, thick] (0,.5ex) -- (0.3,.5ex);} represents the reconstructed one.}
\label{VAWT_AE_power}
\end{figure}

\begin{figure}[hbt!]
\centering
\begin{subfigure}[b]{0.49\textwidth}
    \includegraphics[width=\textwidth]{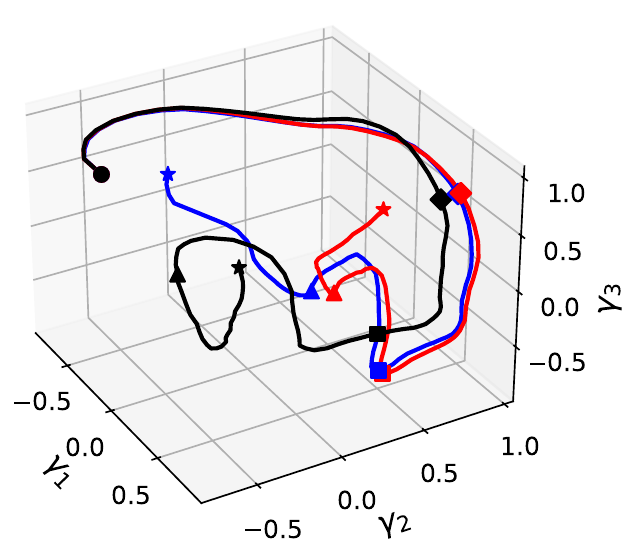}
    \caption{3D visualization of the latent variables trajectories.}
    \label{VAWT_AE_latent_3d}
\end{subfigure}
\hfill 
\begin{subfigure}[b]{0.49\textwidth}
    \includegraphics[width=\textwidth]{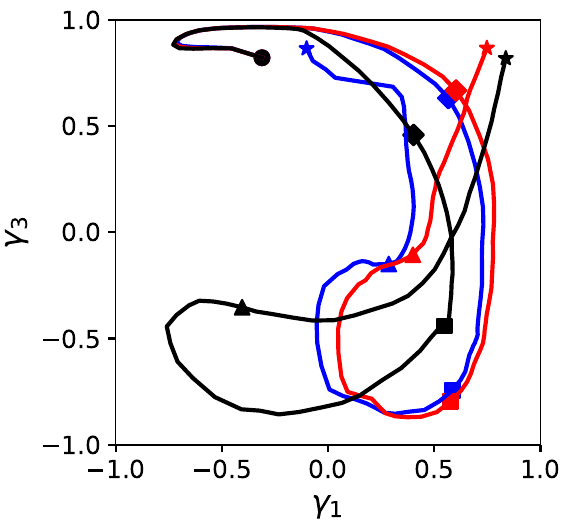}
    \caption{The projection of the trajectories in \fig~\ref{VAWT_AE_latent_3d} on the ($\boldsymbol{\gamma_1}$, $\boldsymbol{\gamma_3}$) plane.}
    
    \label{VAWT_AE_latent_2d}
\end{subfigure}
\caption{Visualization of the latent variables of the same three cases with the same convention of color and symbol use as \fig~\ref{VAWT_AE_vorticity}.}
\label{VAWT_AE_latent}
\end{figure}

The latent variable trajectories corresponding to these cases are illustrated in \fig~\ref{VAWT_AE_latent}. We observe that the trajectories for the first quarter period of all cases are closely aligned. This alignment is expected, given the minimal variations in the vorticity fields and power coefficients across different cases, attributable to the small amplitude of $\ddot{\alpha}$. Further analysis in \fig~\ref{VAWT_AE_latent_3d} reveals that the same portion of these trajectories approximate a circular path when the pitching angle remains relatively small, indicating that the latent space effectively captures essential information about the rotational motion of the flat plate. Additionally, as depicted in \fig~\ref{VAWT_AE_latent_2d}, the latent variable $\gamma_3$ appears to encode information about the rotation angle $\theta$, evidenced by the proximity of the start and end point of all trajectories in the $\gamma_3$ direction---both corresponding to a rotation angle of $0^\circ$---suggesting a consistency in $\gamma_3$ values at these positions. However, as discussed in Section~\ref{sec:pitch_distb_PA-AE}, the precise interpretation and the specific information encoded in the latent space require further exploration in future studies.

\section*{A.3. LDM Training and Validation}
\begin{figure}[hbt!]
\centering
\includegraphics[width=0.8\textwidth]{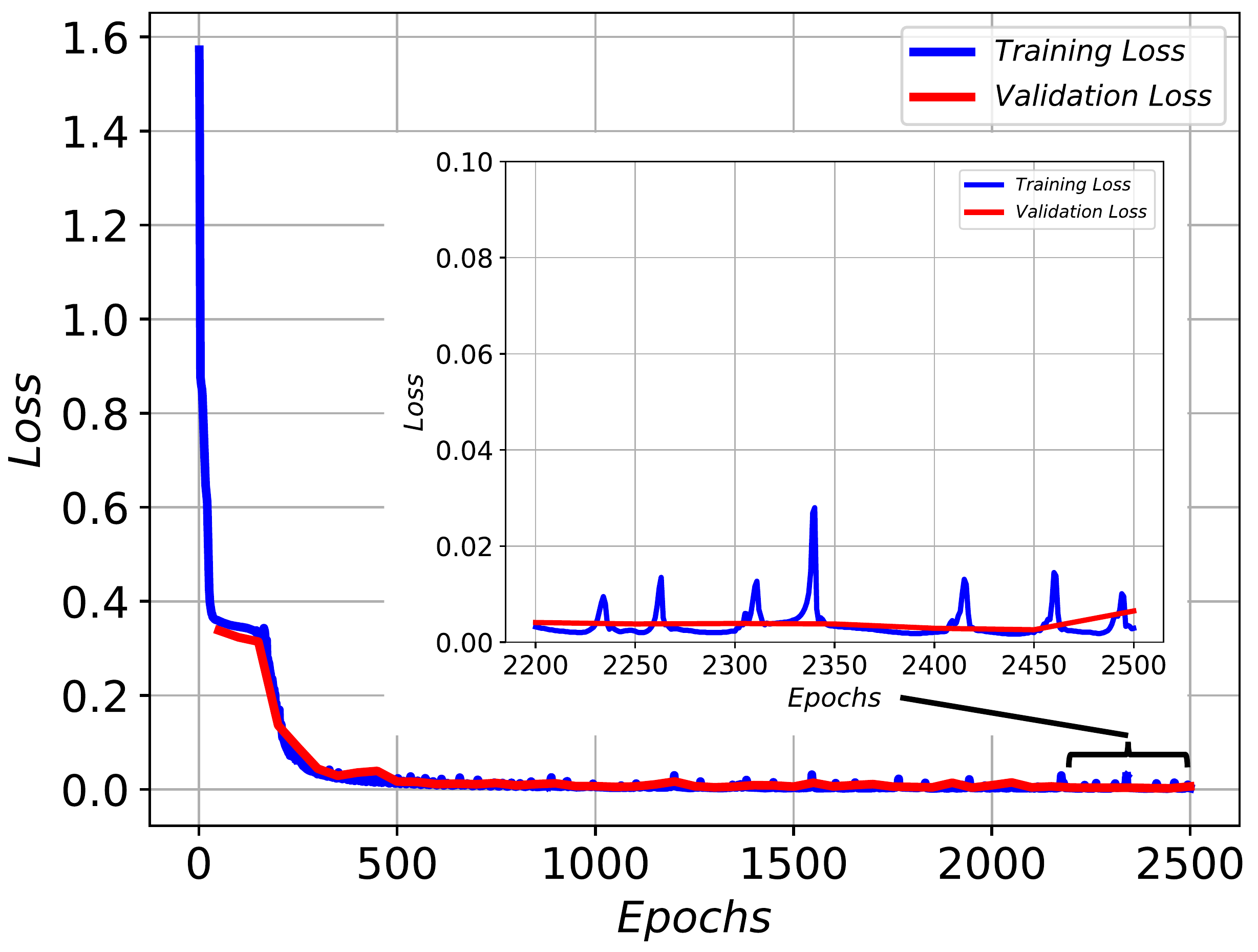}
\caption{The training loss and validation loss history for LDM in VAWT problem.}
\label{VAWT_LSTM_loss}
\end{figure}

After compressing the vorticity field snapshots of dimension of $300\times 300$ to the three-dimensional latent space, we once again utilize the proposed LDM to learn the latent dynamics. In contrast to the previous example in which there was external forcing, in this problem it is sufficient to uniquely learn the dynamics from the initial latent variable $\boldsymbol{\gamma}_1$ and the action sequence $\{A_1,A_2,...,A_k\}$. Therefore, providing the initial condition and the action sequence can uniquely determine how the flow field would change. Consequently, we rearrange the $\boldsymbol{\gamma}_1$ and $\{A_1,A_2,...,A_k\}$ into the input for the LDM, according to condition 1 of \revision{Eq.}~\eqref{eqn:LDM_input}. The output is still defined as \revision{Eq.}~\eqref{eqn:LDM_output}, with the physical variable defined here as the power coefficient. Once again, a standard mean-squared error loss function is used, and to ensure the model generalizes well, we randomly split the $415$ sequences into $315$ sequences for the training dataset, and leave the remaining $100$ sequences for the validation dataset. The corresponding loss histories during the LDM training are shown in \fig~\ref{VAWT_LSTM_loss}. For reference, the LDM training takes approximately $30$ minutes on an NVIDIA GTX $3070$ Ti GPU card.

\begin{figure}[hbt!]
\centering
\includegraphics[width=1.0\textwidth]{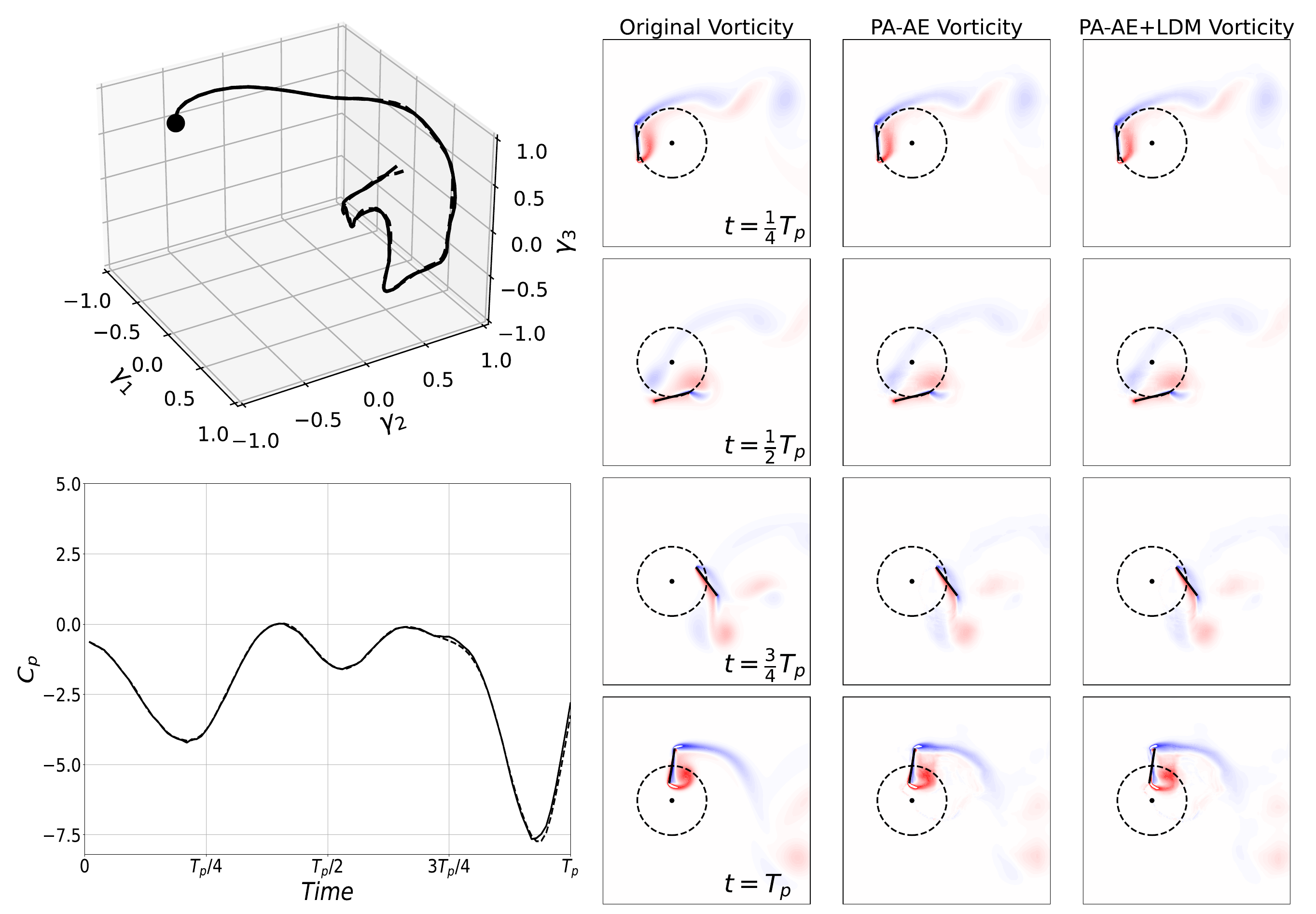}
\caption{The latent variables prediction, power coefficient response prediction, and visualization of the vorticity prediction for unseen case 1. The solid line \protect\tikz[baseline]{\protect\draw[solid, thick] (0,.8ex) -- (0.3,.8ex);} represents the truth, and the dashed line \protect\tikz[baseline]{\protect\draw[dashed, thick] (0,.5ex) -- (0.3,.5ex);} represents the predicted one from LDM for both the latent variable trajectory and the power coefficient history. The vorticity field comparisons among the truth, PA-AE reconstructed one, and PA-AE + LDM predicted one for each quarter of a period are given.}
\label{VAWT_LSTM_case1}
\end{figure}

\begin{figure}[hbt!]
\centering
\includegraphics[width=1.0\textwidth]{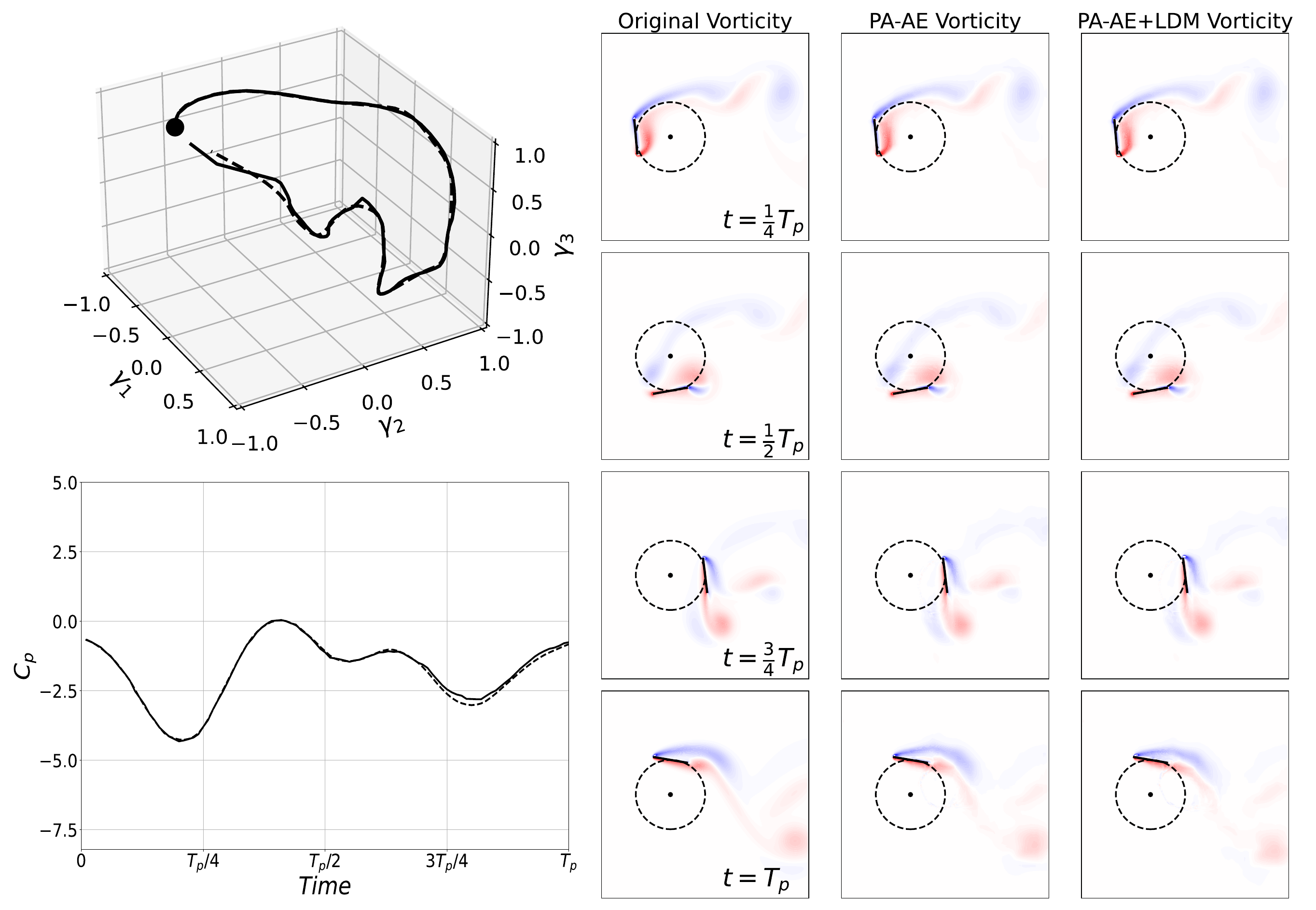}
\caption{The latent variables trajectory, power coefficient response history, and visualization of the vorticity field for unseen case 2. The solid line \protect\tikz[baseline]{\protect\draw[solid, thick] (0,.8ex) -- (0.3,.8ex);} represents the truth, and the dashed line \protect\tikz[baseline]{\protect\draw[dashed, thick] (0,.5ex) -- (0.3,.5ex);} represents the predicted one from LDM for both the latent variable trajectory and the power coefficient history. The vorticity field comparisons among the truth, PA-AE reconstructed one, and PA-AE + LDM predicted one for each quarter of a period are given.}
\label{VAWT_LSTM_case2}
\end{figure}

In \figs~\ref{VAWT_LSTM_case1} and \ref{VAWT_LSTM_case2} we visualize the prediction of the vorticity fields, the latent variable trajectories, and the power coefficient for two unseen cases. First, it is evident that the LDM effectively predicts the latent dynamics throughout the trajectory, maintaining accuracy even as it approaches the end of the rotation. This is also true for the prediction of the power coefficient. Furthermore, the vorticity fields decoded from the latent variables $\boldsymbol{\hat{\gamma}}_k$ predicted by the LDM demonstrate the capacity of the modeled latent space dynamics to encapsulate all crucial aspects of the flow dynamics, including the wake, boundary layer structure, and flow separation phenomena. The high predictive accuracy observed in this problem demonstrates that the proposed LDM can be effectively applied to a wide range of scenarios, including the prediction of flow dynamics associated with complex body motions.

\section*{Appendix \revision{B: PA-AE parameters and the stacked LSTM structure}}
For brevity, the specific architecture \revision{parameters} of the PA-AE in this study \revision{are} not accommodated in the main text, so it is provided in the Tables~\ref{tab:PA-AE parameters pitch case} and \ref{tab:PA-AE parameters VAWT case} for the two scenarios we considered, respectively. The number of convolutional kernel $N_C$, the convolutional kernel size $K_C$, the stride $S_C$, and the amount of padding $P_C$ for the convolutional layer are represented as Conv2D$(N_C,K_C,S_C,P_C)$. The maxpool kernel size $K_M$, the stride $S_M$, and the amoount of padding $P_M$ are represented as MaxPool2D$(K_M,S_M,P_M)$. The scale factor of Upsampling $F_U$, and the corresponding ``nearest'' algorithm are represented as Upsample$(F_U,``\text{nearest}")$.
\setcounter{table}{0}  
\renewcommand{\thetable}{B\arabic{table}} 

\begin{table}[hbt!]
\caption{\label{tab:PA-AE parameters pitch case} PA-AE parameters for \revision{the pitching airfoil problem}}
\centering
\begin{tabular}{cccccc}
\hline
\multicolumn{2}{c}{Encoder} & \multicolumn{2}{c}{Decoder} & \multicolumn{2}{c}{MLP} \\\cline{1-2} \cline{3-4} \cline{5-6}
Layer & Data Size & Layer & Data Size & Layer & Data Size
\\\hline
Input Layer& $(1,240,120)$& Fully Connected& $(16)$& Fully Connected& $(32)$\\
Conv2D$(32,3,1,1$)& $(32,240,120)$& Fully Connected& $(32)$& Fully Connected& $(64)$\\
Conv2D$(32,3,1,1)$& $(32,240,120)$& Fully Connected& $(64)$& Fully Connected& $(32)$\\
MaxPool2D$(2,2,0)$& $(32,120,60)$& Fully Connected& $(256)$& Output& $(1)$\\
Conv2D$(16,3,1,1)$& $(16,120,60)$& Fully Connected& $(288)$& & \\
Conv2D$(16,3,1,1)$& $(16,120,60)$& Reshape& $(4,12,6)$& &\\
MaxPool2D$(2,2,0)$& $(16,60,30)$& Conv2D$(4,3,1,1)$& $(4,12,6)$& & \\
Conv2D$(8,3,1,1)$& $(8,60,30)$& Conv2D$(4,3,1,1)$& $(4,12,6)$& & \\
Conv2D$(8,3,1,1)$& $(8,60,30)$& Upsample$(5,``\text{nearest}")$& (4,60,30)& & \\
MaxPool2D$(5,5,0)$& $(8,12,6)$& Conv2D$(8,3,1,1)$& $(8,60,30)$& & \\
Conv2D$(4,3,1,1)$& $(4,12,6)$& Conv2D$(8,3,1,1)$& $(8,60,30)$& & \\
Conv2D$(4,3,1,1$)& $(4,12,6)$& Upsample$(2,``\text{nearest}")$& $(8,120,60)$& & \\
Reshape & $(288)$& Conv2D$(16,3,1,1)$& $(16,120,60)$& & \\
Fully Connected& $(256)$& Conv2D$(16,3,1,1)$& $(16,120,60)$& & \\
Fully Connected& $(64)$& Upsample$(2,``\text{nearest}")$& $(16,240,120)$& &\\
Fully Connected& $(32)$& Conv2D$(32,3,1,1)$& $(32,240,120)$& & \\
Fully Connected& $(16)$& Conv2D$(32,3,1,1)$& $(32,240,120)$& & \\
Fully Connected& $(3)$& Output Layer& $(1,240,120)$& & \\
\hline
\end{tabular}
\end{table}

\begin{table}[hbt!]
\caption{\label{tab:PA-AE parameters VAWT case} PA-AE parameters for the VAWT problem}
\centering
\begin{tabular}{cccccc}
\hline
\multicolumn{2}{c}{Encoder} & \multicolumn{2}{c}{Decoder} & \multicolumn{2}{c}{MLP} \\\cline{1-2} \cline{3-4} \cline{5-6}
Layer & Data Size & Layer & Data Size & Layer & Data Size
\\\hline
Input Layer& $(1,300,300)$& Fully Connected& $(16)$& Fully Connected& $(32)$\\
Conv2D$(32,3,1,1)$& $(32,300,300)$& Fully Connected& $(32)$& Fully Connected& $(64)$\\
Conv2D$(32,3,1,1)$& $(32,300,300)$& Fully Connected& $(64)$& Fully Connected& $(32)$\\
MaxPool2D$(2,2,0)$& $(32,150,150)$& Fully Connected& $(256)$& Output& $(1)$\\
Conv2D$(16,3,1,1)$& $(16,150,150)$& Fully Connected& $(300)$& & \\
Conv2D$(16,3,1,1)$& $(16,150,150)$& Fully Connected& $(400)$& &\\
MaxPool2D$(3,3,0)$& $(16,50,50)$& Reshape& $(4,10,10)$& & \\
Conv2D$(8,3,1,1)$& $(8,50,50)$& Conv2D$(4,3,1,1)$& $(4,10,10)$& & \\
Conv2D$(8,3,1,1)$& $(8,50,50)$& Conv2D$(4,3,1,1)$& $(4,10,10)$& & \\
MaxPool2D$(5,5,0)$& $(8,10,10)$& Upsample$(5,``\text{nearest}")$& $(4,50,50)$& & \\
Conv2D$(4,3,1,1)$& $(4,10,10)$&Conv2D$(8,3,1,1)$& $(8,50,50)$& & \\
Conv2D$(4,3,1,1)$& $(4,10,10)$& Conv2D$(8,3,1,1)$& $(8,50,50)$& & \\
Reshape & $(400)$& Upsample$(3,``\text{nearest}")$& $(8,150,150)$& & \\
Fully Connected& $(300)$& Conv2D$(16,3,1,1)$& $(16,150,150)$& & \\
Fully Connected& $(256)$& Conv2D$(16,3,1,1)$& $(16,150,150)$& &\\
Fully Connected& $(64)$& Upsample$(2,``\text{nearest}")$& $(16,300,300)$& & \\
Fully Connected& $(32)$& Conv2D$(32,3,1,1)$& $(32,300,300)$& & \\
Fully Connected& $(16)$& Conv2D$(32,3,1,1)$& $(32,300,300)$& & \\
Fully Connected& $(3)$& Output Layer& $(1,300,300)$& & \\
\hline
\end{tabular}
\end{table}

\setcounter{figure}{0}  
\renewcommand{\thefigure}{B\arabic{figure}}  

\begin{figure}[hbt!]
\centering
\begin{subfigure}[b]{1.0\textwidth}
    \includegraphics[width=0.8\textwidth]{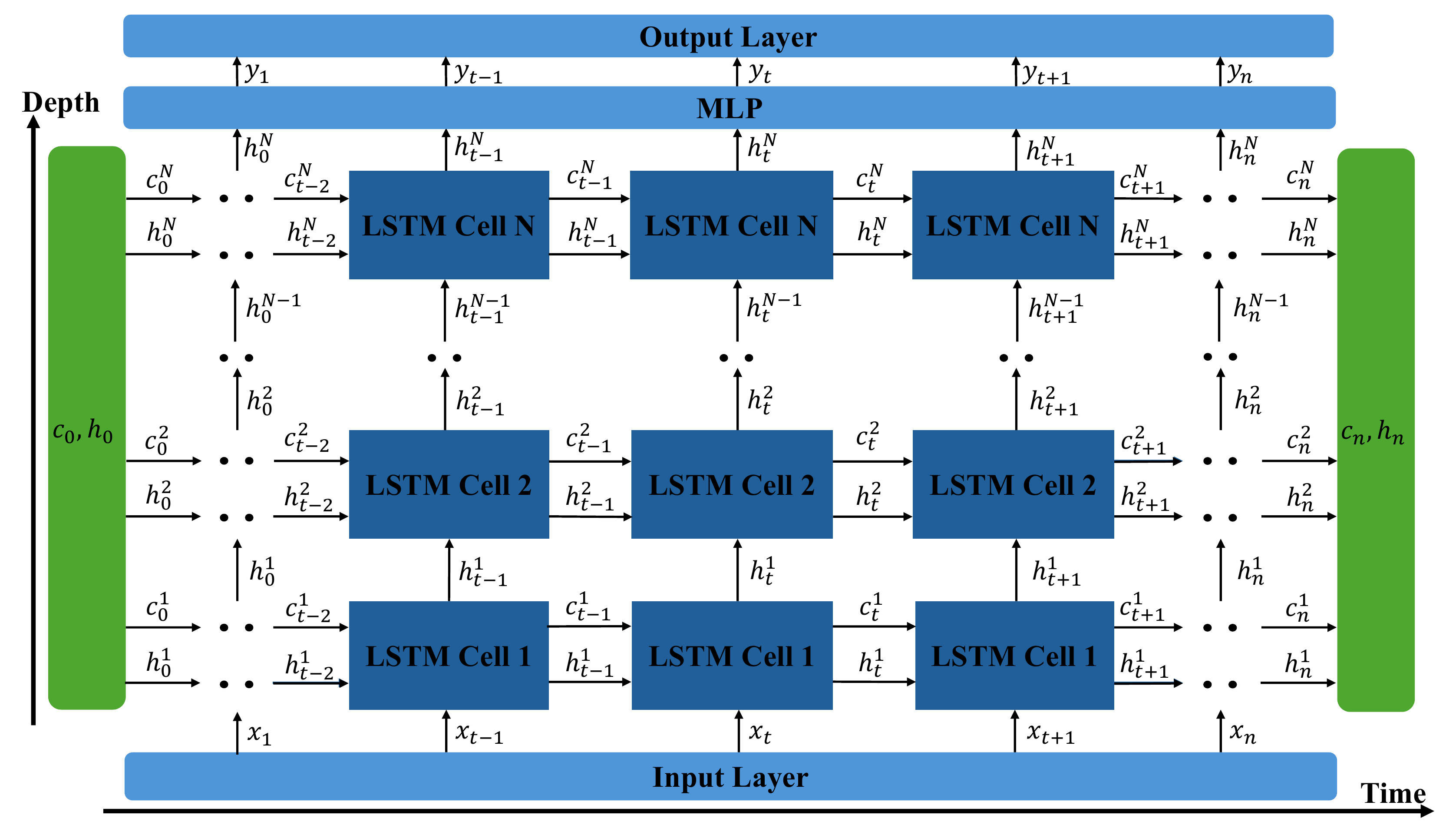}
    \caption{The structure of the stacked LSTM}
    \label{LSTM_overview}
\end{subfigure}
\hfill 
\begin{subfigure}[b]{1.0\textwidth}
    \includegraphics[width=0.8\textwidth]{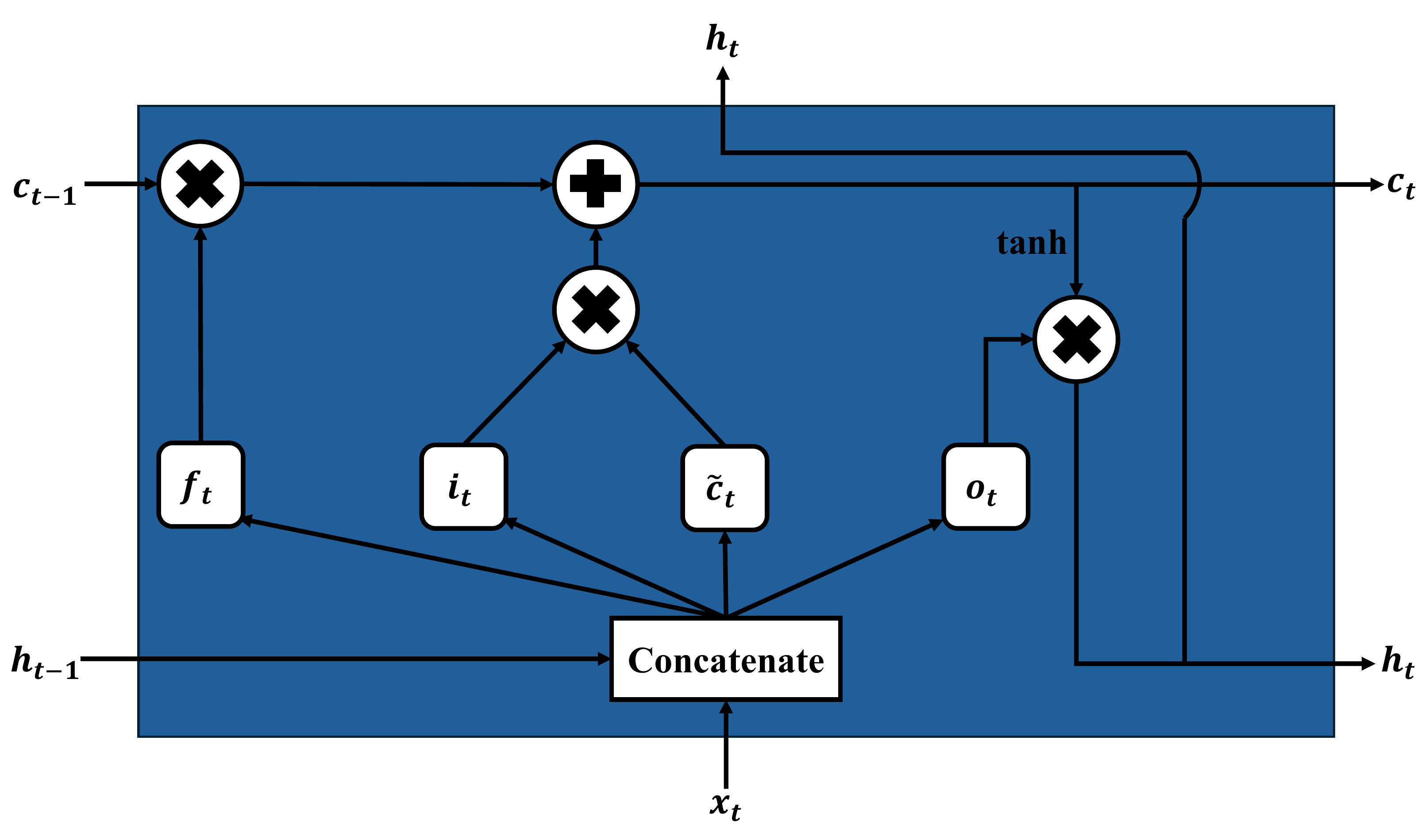}
    \caption{The conceptual structure of the LSTM Cell}
    \label{LSTM_Cell}
\end{subfigure}
\caption{The structure of the stacked LSTM and the LSTM Cell}
\label{LSTM}
\end{figure}

The LDM comprises multiple layers of LSTM cells, each structured to deeply process the temporal sequence data. As depicted in \fig~\ref{LSTM_overview}, each layer, denoted by the superscript \( j \), not only receives information from the output of the preceding layer at the current time step, \( \mathbf{h}_t^{j-1} \), but also carries forward the hidden state and cell state from the previous time step, \( \mathbf{h}_{t-1}^j \) and \( \mathbf{c}_{t-1}^j \), respectively. The output of the final LSTM layer is passed to a MLP, which transforms the high-dimensional hidden states $\{\mathbf{h}_0^N,\mathbf{h}_1^N,...,\mathbf{h}_{n-2}^N\}$ to desired output sequence $\{\mathbf{y}_1,\mathbf{y}_2,...,\mathbf{y}_{n-1}\}$. In this study, the stacked LSTM consists of five layers with hidden state size $32,64,128,64,32$ for each respective layer. After the final LSTM layer, a MLP with two layers, having size $32$ and $4$, respectively, is applied as well. This architecture facilitates the preservation and integration of temporal information across the network layers.

Each LSTM cell across the model’s depth, with structure shown in \fig~\ref{LSTM_Cell}, processes the input sequentially through time, maintaining a memory of past information in its cell state, which helps overcome challenges associated with gradient exploding or vanishing in traditional RNNs, by introducing the forget gate $f_t = \sigma(\mathbf{W}_f \cdot \mathbf{z}_t + \mathbf{b}_f)$ which decides what information to discard from the cell state, input gate $i_t = \sigma(\mathbf{W}_i \cdot \mathbf{z}_t + \mathbf{b}_i)$ which updates the cell state with new information from the current input, output gate $o_t = \sigma(\mathbf{W}_o \cdot \mathbf{z}_t + \mathbf{b}_o)$ which determines the next hidden state based on the updated cell state, and a cell candidate state $\tilde{c}_t = \tanh(\mathbf{W}_c \cdot \mathbf{z}_t + \mathbf{b}_c)$ which generates a candidate vector for updating the cell state, where the $\mathbf{W}_f$, $\mathbf{W}_i$, $\mathbf{W}_o$, $\mathbf{W}_c \in \mathcal{R}^{D_h \times (D_d+D_h)}$, $\mathbf{b}_f$, $\mathbf{b}_i$, $\mathbf{b}_o$, $\mathbf{b}_c \in \mathcal{R}^{(D_d+d_h)}$, respectively, are the weight matrices and bias vectors corresponding to each gate,  and where $D_d, D_h$ represent the dimension of the input and the hidden state, respectively. In addition, $\sigma()$ and $\tanh()$ are the sigmoid and the hyperbolic tangent activation function, respectively. Note that, before the operation of the multiplication with the weight matrices, the LSTM cell would first do a concatenation operation along the column axis such that $\text{Concatenate}(\mathbf{x}_t,\mathbf{h}_{t-1})=\mathbf{z}_t$.

\section*{Acknowledgments}
The authors gratefully acknowledge the financial support for this work provided by the National Science Foundation under Award number 2247005.

\bibliography{main}

\begin{thebibliography}{31}
\newcommand{\enquote}[1]{``#1''}
\providecommand{\natexlab}[1]{#1}
\providecommand{\url}[1]{\texttt{#1}}
\providecommand{\urlprefix}{URL }
\expandafter\ifx\csname urlstyle\endcsname\relax
  \providecommand{\doi}[1]{\discretionary{}{}{}https://doi.org/#1}\else
  \providecommand{\doi}[1]{\discretionary{}{}{}\urlstyle{rm}\url{https://doi.org/#1}}\fi

\bibitem[{Jones et~al.(2022)Jones, Cetiner, and
  Smith}]{annurev:/content/journals/10.1146/annurev-fluid-031621-085520}
Jones, A.~R., Cetiner, O., and Smith, M.~J., \enquote{Physics and Modeling of
  Large Flow Disturbances: Discrete Gust Encounters for Modern Air Vehicles,}
  \emph{Annual Review of Fluid Mechanics}, Vol.~54, No. Volume 54, 2022, 2022,
  pp. 469--493.
\newblock \doi{https://doi.org/10.1146/annurev-fluid-031621-085520},
  \urlprefix\url{https://www.annualreviews.org/content/journals/10.1146/annurev-fluid-031621-085520}.

\bibitem[{{\"O}zkan(2022)}]{Ozkan2022}
{\"O}zkan, M., \emph{Active and Passive Flow Control Methods Over Airfoils for
  Improvement in Aerodynamic Performance}, Springer International Publishing,
  Cham, 2022, pp. 19--33.
\newblock \doi{10.1007/978-3-030-80779-5_2},
  \urlprefix\url{https://doi.org/10.1007/978-3-030-80779-5_2}.

\bibitem[{Ahuja and Rowley(2010)}]{Ahuja2009FeedbackCO}
Ahuja, S., and Rowley, C.~W., \enquote{Feedback control of unstable steady
  states of flow past a flat plate using reduced-order estimators,}
  \emph{Journal of Fluid Mechanics}, Vol. 645, 2010, p. 447–478.
\newblock \doi{10.1017/S0022112009992655}.

\bibitem[{Sedky et~al.(2022)Sedky, Gementzopoulos, Andreu-Angulo, Lagor, and
  Jones}]{Sedky_Gementzopoulos_Andreu-Angulo_Lagor_Jones_2022}
Sedky, G., Gementzopoulos, A., Andreu-Angulo, I., Lagor, F.~D., and Jones,
  A.~R., \enquote{Physics of gust response mitigation in open-loop pitching
  manoeuvres,} \emph{Journal of Fluid Mechanics}, Vol. 944, 2022, p. A38.
\newblock \doi{10.1017/jfm.2022.509}.

\bibitem[{Kerstens et~al.(2011)Kerstens, Pfeiffer, Williams, King, and
  Colonius}]{doi:10.2514/1.J050954}
Kerstens, W., Pfeiffer, J., Williams, D., King, R., and Colonius, T.,
  \enquote{Closed-Loop Control of Lift for Longitudinal Gust Suppression at Low
  Reynolds Numbers,} \emph{AIAA Journal}, Vol.~49, No.~8, 2011, pp. 1721--1728.
\newblock \doi{10.2514/1.J050954},
  \urlprefix\url{https://doi.org/10.2514/1.J050954}.

\bibitem[{Novati et~al.(2017)Novati, Verma, Alexeev, Rossinelli, van Rees, and
  Koumoutsakos}]{Novati_2017}
Novati, G., Verma, S., Alexeev, D., Rossinelli, D., van Rees, W.~M., and
  Koumoutsakos, P., \enquote{Synchronisation through learning for two
  self-propelled swimmers,} \emph{Bioinspiration \& Biomimetics}, Vol.~12,
  No.~3, 2017, p. 036001.
\newblock \doi{10.1088/1748-3190/aa6311},
  \urlprefix\url{https://dx.doi.org/10.1088/1748-3190/aa6311}.

\bibitem[{Verma et~al.(2018)Verma, Novati, and
  Koumoutsakos}]{Verma2018EfficientCS}
Verma, S., Novati, G., and Koumoutsakos, P., \enquote{Efficient collective
  swimming by harnessing vortices through deep reinforcement learning,}
  \emph{Proceedings of the National Academy of Sciences}, Vol. 115, No.~23,
  2018, pp. 5849--5854.
\newblock \doi{10.1073/pnas.1800923115},
  \urlprefix\url{https://www.pnas.org/doi/abs/10.1073/pnas.1800923115}.

\bibitem[{Fan et~al.(2020)Fan, Yang, Wang, Triantafyllou, and
  Karniadakis}]{doi:10.1073/pnas.2004939117}
Fan, D., Yang, L., Wang, Z., Triantafyllou, M.~S., and Karniadakis, G.~E.,
  \enquote{Reinforcement learning for bluff body active flow control in
  experiments and simulations,} \emph{Proceedings of the National Academy of
  Sciences}, Vol. 117, No.~42, 2020, pp. 26091--26098.
\newblock \doi{10.1073/pnas.2004939117},
  \urlprefix\url{https://www.pnas.org/doi/abs/10.1073/pnas.2004939117}.

\bibitem[{Rabault et~al.(2019)Rabault, Kuchta, Jensen, Réglade, and
  Cerardi}]{Rabault_Kuchta_Jensen_Réglade_Cerardi_2019}
Rabault, J., Kuchta, M., Jensen, A., Réglade, U., and Cerardi, N.,
  \enquote{Artificial neural networks trained through deep reinforcement
  learning discover control strategies for active flow control,} \emph{Journal
  of Fluid Mechanics}, Vol. 865, 2019, p. 281–302.
\newblock \doi{10.1017/jfm.2019.62}.

\bibitem[{Varela et~al.(2022)Varela, Suárez, Alcántara-Ávila, Miró,
  Rabault, Font, García-Cuevas, Lehmkuhl, and Vinuesa}]{act11120359}
Varela, P., Suárez, P., Alcántara-Ávila, F., Miró, A., Rabault, J., Font,
  B., García-Cuevas, L.~M., Lehmkuhl, O., and Vinuesa, R., \enquote{Deep
  Reinforcement Learning for Flow Control Exploits Different Physics for
  Increasing Reynolds Number Regimes,} \emph{Actuators}, Vol.~11, No.~12, 2022.
\newblock \doi{10.3390/act11120359},
  \urlprefix\url{https://www.mdpi.com/2076-0825/11/12/359}.

\bibitem[{Beckers and
  Eldredge(2024)}]{beckers2024deepreinforcementlearningairfoil}
Beckers, D., and Eldredge, J.~D., \enquote{Deep reinforcement learning of
  airfoil pitch control in a highly disturbed environment using partial
  observations,} \emph{Phys. Rev. Fluids}, Vol.~9, 2024, p. 093902.
\newblock \doi{10.1103/PhysRevFluids.9.093902},
  \urlprefix\url{https://link.aps.org/doi/10.1103/PhysRevFluids.9.093902}.

\bibitem[{Moerland et~al.(2023)Moerland, Broekens, Plaat, and
  Jonker}]{moerland2023model}
Moerland, T.~M., Broekens, J., Plaat, A., and Jonker, C.~M.,
  \enquote{Model-based Reinforcement Learning: A Survey,} \emph{Foundations and
  Trends® in Machine Learning}, Vol.~16, No.~1, 2023, pp. 1--118.
\newblock \doi{10.1561/2200000086},
  \urlprefix\url{http://dx.doi.org/10.1561/2200000086}.

\bibitem[{Lesort et~al.(2018)Lesort, Díaz-Rodríguez, Goudou, and
  Filliat}]{LESORT2018379}
Lesort, T., Díaz-Rodríguez, N., Goudou, J.-F., and Filliat, D.,
  \enquote{State representation learning for control: An overview,}
  \emph{Neural Networks}, Vol. 108, 2018, pp. 379--392.
\newblock \doi{https://doi.org/10.1016/j.neunet.2018.07.006},
  \urlprefix\url{https://www.sciencedirect.com/science/article/pii/S0893608018302053}.

\bibitem[{Berkooz et~al.(1993)Berkooz, Holmes, and Lumley}]{Berkooz1993ThePO}
Berkooz, G., Holmes, P., and Lumley, J.~L., \enquote{The Proper Orthogonal
  Decomposition in the Analysis of Turbulent Flows,} \emph{Annual Review of
  Fluid Mechanics}, Vol.~25, No. Volume 25, 1993, 1993, pp. 539--575.
\newblock \doi{https://doi.org/10.1146/annurev.fl.25.010193.002543},
  \urlprefix\url{https://www.annualreviews.org/content/journals/10.1146/annurev.fl.25.010193.002543}.

\bibitem[{Taira et~al.(2017)Taira, Brunton, Dawson, Rowley, Colonius, McKeon,
  Schmidt, Gordeyev, Theofilis, and Ukeiley}]{2017_Taira}
Taira, K., Brunton, S.~L., Dawson, S.~A., Rowley, C.~W., Colonius, T., McKeon,
  B.~J., Schmidt, O.~G., Gordeyev, S., Theofilis, V., and Ukeiley, L.,
  \enquote{Modal Analysis of Fluid Flows: An Overview,} \emph{AIAA Journal},
  Vol.~55, No.~12, 2017, pp. 4013--4041.
\newblock \doi{10.2514/1.J056060},
  \urlprefix\url{https://doi.org/10.2514/1.J056060}.

\bibitem[{Hinton and Salakhutdinov(2006)}]{doi:10.1126/science.1127647}
Hinton, G.~E., and Salakhutdinov, R.~R., \enquote{Reducing the Dimensionality
  of Data with Neural Networks,} \emph{Science}, Vol. 313, No. 5786, 2006, pp.
  504--507.
\newblock \doi{10.1126/science.1127647},
  \urlprefix\url{https://www.science.org/doi/abs/10.1126/science.1127647}.

\bibitem[{Murata et~al.(2020)Murata, Fukami, and
  Fukagata}]{Murata_Fukami_Fukagata_2020}
Murata, T., Fukami, K., and Fukagata, K., \enquote{Nonlinear mode decomposition
  with convolutional neural networks for fluid dynamics,} \emph{Journal of
  Fluid Mechanics}, Vol. 882, 2020, p. A13.
\newblock \doi{10.1017/jfm.2019.822}.

\bibitem[{Fukami and Taira(2023)}]{Fukami2023}
Fukami, K., and Taira, K., \enquote{Grasping extreme aerodynamics on a
  low-dimensional manifold,} \emph{Nature Communications}, Vol.~14, No.~1,
  2023, p. 6480.
\newblock \doi{10.1038/s41467-023-42213-6},
  \urlprefix\url{https://doi.org/10.1038/s41467-023-42213-6}.

\bibitem[{Schmid(2010)}]{SCHMID_2010}
Schmid, P.~J., \enquote{Dynamic mode decomposition of numerical and
  experimental data,} \emph{Journal of Fluid Mechanics}, Vol. 656, 2010, p.
  5–28.
\newblock \doi{10.1017/S0022112010001217}.

\bibitem[{Brunton et~al.(2016)Brunton, Proctor, and
  Kutz}]{doi:10.1073/pnas.1517384113}
Brunton, S.~L., Proctor, J.~L., and Kutz, J.~N., \enquote{Discovering governing
  equations from data by sparse identification of nonlinear dynamical systems,}
  \emph{Proceedings of the National Academy of Sciences}, Vol. 113, No.~15,
  2016, pp. 3932--3937.
\newblock \doi{10.1073/pnas.1517384113},
  \urlprefix\url{https://www.pnas.org/doi/abs/10.1073/pnas.1517384113}.

\bibitem[{Linot et~al.(2023{\natexlab{a}})Linot, Burby, Tang, Balaprakash,
  Graham, and Maulik}]{LINOT2023111838}
Linot, A.~J., Burby, J.~W., Tang, Q., Balaprakash, P., Graham, M.~D., and
  Maulik, R., \enquote{Stabilized neural ordinary differential equations for
  long-time forecasting of dynamical systems,} \emph{Journal of Computational
  Physics}, Vol. 474, 2023{\natexlab{a}}, p. 111838.
\newblock \doi{https://doi.org/10.1016/j.jcp.2022.111838},
  \urlprefix\url{https://www.sciencedirect.com/science/article/pii/S0021999122009019}.

\bibitem[{Linot et~al.(2023{\natexlab{b}})Linot, Zeng, and
  Graham}]{linot2023turbulence}
Linot, A.~J., Zeng, K., and Graham, M.~D., \enquote{Turbulence control in plane
  Couette flow using low-dimensional neural ODE-based models and deep
  reinforcement learning,} \emph{International Journal of Heat and Fluid Flow},
  Vol. 101, 2023{\natexlab{b}}, p. 109139.
\newblock \doi{https://doi.org/10.1016/j.ijheatfluidflow.2023.109139},
  \urlprefix\url{https://www.sciencedirect.com/science/article/pii/S0142727X23000383}.

\bibitem[{Zeng et~al.(2022)Zeng, Linot, and
  Graham}]{doi:10.1098/rspa.2022.0297}
Zeng, K., Linot, A.~J., and Graham, M.~D., \enquote{Data-driven control of
  spatiotemporal chaos with reduced-order neural ODE-based models and
  reinforcement learning,} \emph{Proceedings of the Royal Society A:
  Mathematical, Physical and Engineering Sciences}, Vol. 478, No. 2267, 2022,
  p. 20220297.
\newblock \doi{10.1098/rspa.2022.0297},
  \urlprefix\url{https://royalsocietypublishing.org/doi/abs/10.1098/rspa.2022.0297}.

\bibitem[{Hochreiter and Schmidhuber(1997)}]{10.1162/neco.1997.9.8.1735}
Hochreiter, S., and Schmidhuber, J., \enquote{{Long Short-Term Memory},}
  \emph{Neural Computation}, Vol.~9, No.~8, 1997, pp. 1735--1780.
\newblock \doi{10.1162/neco.1997.9.8.1735},
  \urlprefix\url{https://doi.org/10.1162/neco.1997.9.8.1735}.

\bibitem[{Brunton et~al.(2020)Brunton, Noack, and
  Koumoutsakos}]{annurev:/content/journals/10.1146/annurev-fluid-010719-060214}
Brunton, S.~L., Noack, B.~R., and Koumoutsakos, P., \enquote{Machine Learning
  for Fluid Mechanics,} \emph{Annual Review of Fluid Mechanics}, Vol.~52, No.
  Volume 52, 2020, 2020, pp. 477--508.
\newblock \doi{https://doi.org/10.1146/annurev-fluid-010719-060214},
  \urlprefix\url{https://www.annualreviews.org/content/journals/10.1146/annurev-fluid-010719-060214}.

\bibitem[{Vlachas et~al.(2018)Vlachas, Byeon, Wan, Sapsis, and
  Koumoutsakos}]{doi:10.1098/rspa.2017.0844}
Vlachas, P.~R., Byeon, W., Wan, Z.~Y., Sapsis, T.~P., and Koumoutsakos, P.,
  \enquote{Data-driven forecasting of high-dimensional chaotic systems with
  long short-term memory networks,} \emph{Proceedings of the Royal Society A:
  Mathematical, Physical and Engineering Sciences}, Vol. 474, No. 2213, 2018,
  p. 20170844.
\newblock \doi{10.1098/rspa.2017.0844},
  \urlprefix\url{https://royalsocietypublishing.org/doi/abs/10.1098/rspa.2017.0844}.

\bibitem[{Vlachas et~al.(2020)Vlachas, Pathak, Hunt, Sapsis, Girvan, Ott, and
  Koumoutsakos}]{VLACHAS2020191}
Vlachas, P., Pathak, J., Hunt, B., Sapsis, T., Girvan, M., Ott, E., and
  Koumoutsakos, P., \enquote{Backpropagation algorithms and Reservoir Computing
  in Recurrent Neural Networks for the forecasting of complex spatiotemporal
  dynamics,} \emph{Neural Networks}, Vol. 126, 2020, pp. 191--217.
\newblock \doi{https://doi.org/10.1016/j.neunet.2020.02.016},
  \urlprefix\url{https://www.sciencedirect.com/science/article/pii/S0893608020300708}.

\bibitem[{Fujimoto et~al.(2018)Fujimoto, van Hoof, and
  Meger}]{DBLP:journals/corr/abs-1802-09477}
Fujimoto, S., van Hoof, H., and Meger, D., \enquote{Addressing Function
  Approximation Error in Actor-Critic Methods,} \emph{CoRR}, Vol.
  abs/1802.09477, 2018.
\newblock \doi{https://doi.org/10.48550/arXiv.1802.09477},
  \urlprefix\url{http://arxiv.org/abs/1802.09477}.

\bibitem[{Eldredge(2022)}]{eldredge2022method}
Eldredge, J.~D., \enquote{A method of immersed layers on Cartesian grids, with
  application to incompressible flows,} \emph{Journal of Computational
  Physics}, Vol. 448, 2022, p. 110716.
\newblock \doi{https://doi.org/10.1016/j.jcp.2021.110716},
  \urlprefix\url{https://www.sciencedirect.com/science/article/pii/S0021999121006112}.

\bibitem[{Raffin et~al.(2021)Raffin, Hill, Gleave, Kanervisto, Ernestus, and
  Dormann}]{JMLR:v22:20-1364}
Raffin, A., Hill, A., Gleave, A., Kanervisto, A., Ernestus, M., and Dormann,
  N., \enquote{Stable-Baselines3: Reliable Reinforcement Learning
  Implementations,} \emph{Journal of Machine Learning Research}, Vol.~22, No.
  268, 2021, pp. 1--8.
\newblock \urlprefix\url{http://jmlr.org/papers/v22/20-1364.html}.

\bibitem[{Le~Fouest and Mulleners(2024)}]{le2024optimal}
Le~Fouest, S., and Mulleners, K., \enquote{Optimal blade pitch control for
  enhanced vertical-axis wind turbine performance,} \emph{Nature
  Communications}, Vol.~15, No.~1, 2024, p. 2770.
\newblock \doi{https://doi.org/10.1038/s41467-024-46988-0}.

\end{thebibliography}

\end{document}